\documentclass[aps,pra,amsmath,two column,amssymb,showpacs]{revtex4}
\usepackage{txfonts}
\usepackage{mathbbold}
\usepackage{epsfig,bm,dcolumn}
\usepackage{graphicx}
\usepackage{color}
\usepackage{overpic}

\begin{document}

\title{Quantum fluctuations in the BCS-BEC crossover of two-dimensional Fermi gases}

\author{Lianyi He$^{1}$, Haifeng L\"u$^{2}$, Gaoqing Cao$^{3}$, Hui Hu$^{4}$, and Xia-Ji Liu$^{4}$}

\affiliation{1 Theoretical Division, Los Alamos National Laboratory, Los Alamos, New Mexico 87545, USA\\
2 Department of Applied Physics, University of Electronic Science and Technology of China, Chengdu 610054, China\\
3 Department of Physics and Collaborative Innovation Center of Quantum Matter, Tsinghua University, Beijing 100084, China\\
4 Centre for Quantum and Optical Science, Swinburne University of Technology, Melbourne 3122, Australia}

\date{\today}

\begin{abstract}
We present a theoretical study of the ground state of the BCS-BEC crossover in dilute two-dimensional Fermi gases. While the mean-field theory provides a simple and analytical equation of state, the pressure is equal to that of a noninteracting Fermi gas in the entire BCS-BEC crossover, which is not consistent with the features of a weakly interacting Bose condensate in the BEC limit and a weakly interacting Fermi liquid in the BCS limit. The inadequacy of the 2D mean-field theory indicates that the quantum fluctuations are much more pronounced than those in 3D. In this work, we show that the inclusion of the Gaussian quantum fluctuations naturally recovers the above features in both the BEC and the BCS limits. In the BEC limit, the missing logarithmic dependence on the boson chemical potential is recovered by the quantum fluctuations. Near the quantum phase transition from the vacuum to the BEC phase, we compare our equation of state with the known grand canonical equation of state of 2D Bose gases and determine the ratio of the composite boson scattering length
$a_{\rm B}$ to the fermion scattering length $a_{\rm 2D}$. We find $a_{\rm B}\simeq 0.56 a_{\rm 2D}$, in good agreement with the exact four-body calculation. We compare our equation of state in the BCS-BEC crossover with recent results from the quantum Monte Carlo simulations and the experimental measurements and find good agreements.
\end{abstract}


\maketitle

\section{Introduction}\label{s1}

The experimental realization of ultracold atomic Fermi gases with tunable interatomic interactions has opened a new era for the study
of some longstanding theoretical proposals in many-fermion systems.  One interesting proposal is the smooth crossover from a
Bardeen-Cooper-Schrieffer (BCS) superfluid ground state with largely overlapping Cooper pairs to a Bose-Einstein condensate (BEC) of
tightly bound bosonic molecules -- a phenomenon suggested many years ago~\cite{Eagles,Leggett,NSR}. A simple but important system is
a dilute attractive Fermi gas in three dimensions (3D), where the effective range of the short-ranged interaction is much smaller
than the interparticle distance. The system can be characterized by a dimensionless gas parameter $1/(k_{\rm F}a_{\rm 3D})$, where
$a_{\rm 3D}$ is the $s$-wave scattering length of the short-ranged interaction and $k_{\rm F}$ is the Fermi momentum in the absence
of interaction. The BCS-BEC crossover occurs when the parameter $1/(k_{\rm F}a_{\rm 3D})$ is tuned from negative to positive values
~\cite{BCSBEC1,BCSBEC2,BCSBEC3,BCSBEC4,BCSBEC5}, and the BCS and BEC limits correspond to the cases $1/(k_{\rm F}a_{\rm 3D})\rightarrow-\infty$
and $1/(k_{\rm F}a_{\rm 3D})\rightarrow+\infty$, respectively.

The BCS-BEC crossover phenomenon in 3D dilute Fermi gases has been experimentally demonstrated by using ultracold gases of $^6$Li and
$^{40}$K atoms~\cite{BCSBECexp1,BCSBECexp2,BCSBECexp3}, where the $s$-wave scattering length and hence the gas parameter
$1/(k_{\rm F}a_{\rm 3D})$ were tuned by means of the Feshbach resonance~\cite{FR1,FR2}. The equation of state and various static and
dynamic properties of the BCS-BEC crossover have become a big challenge for quantum many-body
theory~\cite{TH01,TH02,TH03,TH04,TH05,TH06,TH07,TH08,TH09,TH10,TH11}
because the conventional perturbation theory is no longer valid. At the so-called unitary point where $a_{\rm 3D}\rightarrow\infty$, the only
length scale of the system is the inter-particle distance. Therefore, the properties of the system at the unitary point $1/(k_{\rm F}a_{\rm 3D})=0$
become universal, i.e., independent of the details of the interactions. All thermodynamic quantities, scaled by their counterparts for the
non-interacting Fermi gases, become universal constants. Determining these universal constants has been one of the most intriguing topics
in the research of the cold Fermi gases~\cite{EOSexp1,EOSexp2,EOSexp3,EOSmc1,EOSmc2,EOSmc3,EOSmc4}.

On the other hand, it was suggested that a 2D Fermi gas with short-ranged $s$-wave attraction can also undergo a BCS-BEC
crossover~\cite{BCSBEC2D-1,BCSBEC2D-2,BCSBEC2D-3}. Unlike 3D, a two-body bound state always exists in 2D even though the attraction is
arbitrarily weak. The BCS-BEC crossover in 2D can be realized by tuning the binding energy of the bound state. Studying the BCS-BEC crossover in 2D will help us understand the physics of pseudogap and Berezinskii-Kosterlitz-Thouless transitions in fermionic systems~\cite{BCSBEC2D-4}.
In recent years, quasi-2D atomic Fermi gases have been experimentally realized and studied by a number of groups~
\cite{2Dexp1,2Dexp2,2Dexp3,2Dexp4,2Dexp5,2Dexp6,2Dexp7,2Dexp8,2Dexp9,2Dexp10}. In cold-atom experiments, a quasi-2D Fermi gas can be
realized by arranging a one-dimensional optical lattice along the axial direction and a weak harmonic trapping potential in the radial
plane, such that fermions are strongly confined along the axial direction and form a series of pancake-shaped quasi-2D clouds. The
strong anisotropy of the trapping potentials, namely, $\omega_{z}\gg \omega_\bot$ where $\omega_z$ ($\omega_\bot$) is the axial (radial)
frequency, allows us to use an effective 2D Hamiltonian to deal with the radial degrees of freedom. Experimental studies of quasi-2D
Fermi gases have promoted great theoretical interests in the past few years
~\cite{QMC2D1,QMC2D2,QMC2D3,2DTH01,2DTH02,2DTH03,2DTH04,2DTH05,2DTH06,2DTH07,2DTH08,2DTHs,2DTH09,
2DTH10,2DTH11,2DTH12,2DTH13,2DTH14,2DTH15,2DTH16,2DTH17,2DTH18,2DTH19,2DTH20,2DTH21,2DTH22,2DTH23,2DTH24,2DTH25}.

It is known that in 3D,  even the mean-field theory predicts that the system is a weakly interacting Bose condensate in the
strong attraction limit~\cite{BCSBEC2}. The composite boson scattering length is shown to be $a_{\rm B}=2a_{\rm 3D}$~\cite{BCSBEC2}. The
inclusion of Gaussian pair fluctuations~\cite{TH03,TH04,TH05} recovers the Fermi liquid corrections in the weak attraction limit and modifies the composite boson scattering length to $a_{\rm B}\simeq0.55a_{\rm 3D}$, which is close to the exact result $a_{\rm B}\simeq0.6a_{\rm 3D}$~\cite{FB3D}.  Moreover, the equation of state (EOS) in the BCS-BEC crossover agrees excellently with the quantum Monte Carlo results and the experimental measurements if the Gaussian pair fluctuations are taken into account~\cite{TH03,TH04,TH05}. In contrast, the mean-field theory for 2D Fermi gases does not predict a weakly interacting 2D Bose condensate in the strong attraction limit~\cite{BCSBEC2D-2,BCSBEC2D-3}. The coupling constant between the composite bosons is predicted to be energy independent, which arises from the inadequacy of the Born approximation for four-body scattering in 2D. As a result, the 2D mean-field theory predicts that the pressure of a homogeneous 2D Fermi gas is equal to that of a noninteracting Fermi gas in the entire BCS-BEC crossover. However, recent experimental measurements~\cite{2Dexp2,2Dexp6} and quantum Monte Carlo simulations~\cite{QMC2D1,QMC2D2,QMC2D3} show that the pressure in the strong attraction limit is vanishingly small in comparison to that of a noninteracting Fermi gas, which is consistent with the picture that the system is a weakly interacting 2D Bose condensate. These results indicate that the 2D mean-field theory is not adequate even at the qualitative level, and quantum fluctuations are much more important in 2D.

In analogy to the 3D case, we expect that the inclusion of Gaussian pair fluctuations in 2D naturally recovers the feature of the weakly interacting 2D Bose condensate in the strong attraction limit. This has been demonstrated recently by using the pole approximation for the Goldstone mode and the dimensional regularization for the untraviolet divergence, which leads to an elegant derivation of the composite boson scattering length~\cite{2DTH24}. However, the pole approximation is limited in the strong attraction limit because of the use of the Bogoliubov dispersion for the Goldstone mode. We also note that the ultraviolet (UV) divergence arising from the pole approximation can be naturally avoided in the full treatment of the collective modes~\cite{TH03,TH04,TH05}. In this work, we study the influence of quantum fluctuations on the EOS of 2D Fermi gases in the entire BCS-BEC crossover. With the full EOS beyond the pole approximation, we determine the ratio of the composite boson scattering length $a_{\rm B}$ to the fermion scattering length $a_{\rm 2D}$ by comparing our EOS near the vacuum-BEC quantum phase transition with the known grand canonical EOS of weakly interacting 2D Bose gases~\cite{2DBEC01,2DBEC02,2DBEC03,2DBEC04,2DBEC05,2DBEC06,2DBEC07}. We obtain $a_{\rm B}\simeq 0.56 a_{\rm 2D}$, in good agreement with the exact four-body calculation~\cite{2DTH01} and the pole approximation treatment \cite{2DTH24}. We also perform numerical calculations for the canonical EOS of a homogeneous 2D Fermi gas in the BCS-BEC crossover. In addition to recovering the weakly interacting Bose condensate in the strong attraction limit, we find that the Fermi liquid corrections~\cite{PEOS2D1,PEOS2D2,PEOS2D3} can also be recovered at sufficiently weak attraction. We compare our EOS with the recent results from quantum Monte Carlo simulations and experimental measurements in the entire BCS-BEC crossover and find good agreements.

The paper is organized as follows. We set up our theoretical framework for 2D Fermi gases beyond mean field in Sec. \ref{s2}. We study the strong attraction limit and determine the composite boson scattering length in Sec. \ref{s3}. We present our theoretical predictions for the EOS in the BCS-BEC crossover and compare our results with the quantum Monte Carlo data and experimental measurements in Sec. \ref{s4}. We summarize in Sec. \ref{s5}. The natural units $\hbar=k_{\text B}=1$ are used throughout.

\section{Hamiltonian and grand potential}\label{s2}

We consider a spin-$1/2$ (two-component) Fermi gas in two spatial dimensions with a short-ranged $s$-wave attractive interaction between
the unlike spins. In the dilute limit, the interaction potential can be safely modeled by a contact interaction. The grand canonical
Hamiltonian density of the system is given by
\begin{eqnarray}
H=\sum_{\sigma=\uparrow,\downarrow}\bar{\psi}_\sigma({\bf r}){\cal H}_0 \psi_\sigma({\bf
r})-U\bar{\psi}_{\uparrow}({\bf
r})\bar{\psi}_{\downarrow}({\bf r})\psi_{\downarrow}({\bf
r})\psi_{\uparrow}({\bf r}),\label{HBCS}
\end{eqnarray}
where $\psi_\uparrow({\bf r})$ and $\psi_\downarrow({\bf r})$ represent the annihilation field operators for the two spin states of fermions,
${\cal H}_0=-\nabla^2/(2m)-\mu$ is the free single-particle Hamiltonian, with $m$ being the fermion mass and $\mu$ being the chemical potential,
and $U>0$ denotes the $s$-wave attractive interaction occurring between unlike spins.

The contact coupling $U$ is convenient for performing theoretical derivations. However, it should be renormalized by using some physical quantities so that we can obtain finite results in the many-body calculations. With the contact interaction $U$, the Lippmann-Schwinger equation
for the two-body $T$ matrix reads
\begin{equation}
T_{\rm 2B}^{-1}(E)=-U^{-1}-\Pi(E),
\end{equation}
where $E=k^2/m$ is the scattering energy in the center-of-mass frame and the two-particle bubble function $\Pi(E)$ is given by
\begin{equation}
\Pi(E)=\sum_{\bf k}\frac{1}{E+i\epsilon-2\varepsilon_{\bf k}}.
\end{equation}
Here $\epsilon=0^+$ and $\varepsilon_{\bf k}={\bf k}^2/(2m)$. We use the notation $\sum_{\bf k}\equiv \int d^2{\bf k}/(2\pi)^2$ throughout. The cost of the use of the contact interaction is that the integral over ${\bf k}$ suffers from UV divergence. We regularize the UV divergence by introducing a hard cutoff $\Lambda$ for $|{\bf k}|$. For large $\Lambda$ we obtain
\begin{equation}
\Pi(E)=-\frac{m}{4\pi}\ln\frac{\Lambda^2}{m}+\frac{m}{4\pi}\ln\left(-E-i\epsilon\right).
\end{equation}
Next we match the scattering amplitude $f(k)=(4\pi/m) T_{\rm 2B}(E)$ to the known 2D $s$-wave scattering amplitude in the zero-range limit, which is given by $f(k)=1/[\ln(\varepsilon_{\rm B}/E)+i\pi]$~\cite{BCSBEC2D-2,BCSBEC2D-3}. Here $\varepsilon_{\rm B}$ is the binding energy of the two-body bound state which characterizes the attraction strength. Thus we obtain
\begin{eqnarray}\label{URG}
\frac{1}{U(\Lambda)}=\frac{m}{4\pi}\ln\frac{\Lambda^2}{m\varepsilon_{\rm B}}=
\sum_{|{\bf k}|<\Lambda}\frac{1}{2\varepsilon_{\bf k}+\varepsilon_{\rm B}}.
\end{eqnarray}
The above results should be understood in the large-$\Lambda$ limit. After the renormalization of the bare coupling $U$ through the physical binding energy $\varepsilon_{\rm B}$, the UV divergence in the many-body calculations can be eliminated and we can set $\Lambda\rightarrow\infty$ to obtain the final finite results.

In the imaginary-time functional path integral formalism, the partition function at temperature $T$ is
\begin{equation}
{\cal Z} = \int [d\psi][d\bar{\psi}]\exp\left\{-{\cal S}[\psi,\bar{\psi}]\right\},
\end{equation}
where the action
\begin{equation}
{\cal S}[\psi,\bar{\psi}]=\int dx\left[\bar{\psi}\partial_\tau \psi+H(\psi,\bar{\psi})\right].
\end{equation}
Here $x=(\tau, {\bf r})$, with $\tau$ being the imaginary time, and $\int dx=\int_0^\beta d\tau\int d^2{\bf r}$ with $\beta=1/T$. To decouple the interaction term, we introduce an auxiliary pairing field $\Phi(x)$ and apply the Hubbard-Stratonovich transformation. Then the the partition function can be expressed as
\begin{eqnarray}
{\cal Z}=\int [d\psi][d\bar{\psi}][d\Phi][d\Phi^*] \exp\Big\{-{\cal S}[\psi,\bar{\psi},\Phi,\Phi^*]\Big\},
\end{eqnarray}
where the action  reads
\begin{eqnarray}
{\cal S}=\int dx\frac{|\Phi(x)|^2}{U}-\int dx\int dx^\prime\bar{\psi}(x){\bf G}^{-1}(x,x^\prime)\psi(x^\prime).
\end{eqnarray}
Here the Nambu-Gor'kov spinor $\psi(x)=[\psi_\uparrow(x),\bar{\psi}_\downarrow(x)]^{\rm T}$ is employed and the inverse Green's function
${\bf G}^{-1}(x,x^\prime)$ in the Nambu-Gor'kov representation is given by
\begin{eqnarray}
{\bf G}^{-1}(x,x^\prime)=\left(\begin{array}{cc}-\partial_{\tau}-{\cal
H}_0 &\Phi(x)\\  \Phi^*(x)& -\partial_{\tau}+{\cal
H}_0\end{array}\right)\delta(x-x^\prime).
\end{eqnarray}
Integrating out the fermion fields, we obtain
\begin{eqnarray}
\mathcal {Z}=\int[d\Phi] [d\Phi^{\ast}] \exp \Big\{- {\cal
S}_{\rm{eff}}[\Phi, \Phi^{\ast}]\Big\},
\end{eqnarray}
where the effective action reads
\begin{equation}
{\cal S}_{\rm{eff}}[\Phi, \Phi^{\ast}] = \frac{1}{U}\int dx\
|\Phi(x)|^{2} - \mbox{Trln} [{\bf G}^{-1}(x,x^\prime)].
\label{effaction}
\end{equation}

The partition function cannot be evaluated analytically since the path integral over $\Phi$ and $\Phi^*$ cannot be carried out.
At $T=0$, the pairing field $\Phi(x)$ acquires a nonzero and uniform expectation value $\langle\Phi(x)\rangle=\Delta$, which
serves as the order parameter of superfluidity. Due to the U$(1)$ symmetry, we can set $\Delta$ to be real and positive without loss of
generality. Then we write $\Phi(x)=\Delta+\phi(x)$, where $\phi(x)$ is the fluctuation around the mean field. The effective action
${\cal S}_{\text{eff}}[\Phi,\Phi^*]$ can be expanded in powers of the fluctuation $\phi(x)$; that is,
\begin{eqnarray}
{\cal S}_{\text{eff}}[\Phi,\Phi^*]={\cal S}_{\rm MF}
+{\cal S}_{\rm GF}[\phi,\phi^*]+\cdots,
\end{eqnarray}
where ${\cal S}_{\rm MF}\equiv{\cal S}_{\rm eff}[\Delta,\Delta]$ is the saddle-point or mean-field effective action and
${\cal S}_{\rm GF}[\phi,\phi^*]$ is the contribution from the Gaussian fluctuations (GFs). The higher-order contributions from non-Gaussian fluctuations are not shown. Accordingly, the grand potential can be expressed as
\begin{equation}
\Omega=\Omega_{\rm MF}+\Omega_{\rm GF}+\cdots,
\end{equation}
where $\Omega_{\rm MF}={\cal S}_{\rm MF}/(\beta V)$, with $V$ being the volume, and $\Omega_{\rm GF}$ is the contribution from
the Gaussian fluctuations.

\subsection{Mean-field approximation}

In the mean-field approximation, the effective action is approximated as ${\cal S}_{\text{eff}}[\Phi,\Phi^*]\simeq{\cal S}_{\rm MF}$.
The quantum fluctuations are completely neglected. At $T=0$, the mean-field grand potential can be evaluated as
\begin{equation}
\Omega_{\rm MF}=\frac{\Delta^2}{U}+\sum_{\bf k}\left(\xi_{\bf k}-E_{\bf k}\right),
\end{equation}
where $\xi_{\bf k}=\varepsilon_{\bf k}-\mu$ and $E_{\bf k}=\sqrt{\xi_{\bf k}^2+\Delta^2}$. The UV divergence can be eliminated by using
Eq. (\ref{URG}). We obtain
\begin{eqnarray}
\Omega_{\rm MF}=\Delta^2\sum_{\bf k}\left(\frac{1}{2\varepsilon_{\bf k}+\varepsilon_{\rm B}}-\frac{1}{E_{\bf k}+\xi_{\bf k}}\right).
\end{eqnarray}
The order parameter $\Delta$ should be determined as a function of $\mu$ by using the extreme condition $\partial\Omega_{\rm MF}/\partial\Delta=0$. We obtain the gap equation
\begin{equation}\label{MFGAPE}
\frac{1}{U}=\sum_{\bf k}\frac{1}{2E_{\bf k}}
\end{equation}
or, explicitly,
\begin{equation}
\sum_{\bf k}\left(\frac{1}{2\varepsilon_{\bf k}+\varepsilon_{\rm B}}-\frac{1}{2E_{\bf k}}\right)=0.
\end{equation}

It is very fortunate that in 2D the integral over ${\bf k}$ can be carried out. The grand potential reads
\begin{eqnarray}
\Omega_{\rm MF}=\frac{m\Delta^2}{4\pi}\left(\ln\frac{\sqrt{\mu^2+\Delta^2}-\mu}{\varepsilon_{\rm B}}
-\frac{\mu}{\sqrt{\mu^2+\Delta^2}-\mu}-\frac{1}{2}\right).
\end{eqnarray}
Using the extreme condition $\partial\Omega_{\rm MF}/\partial\Delta=0$, we obtain
\begin{equation}
\Delta_{\rm MF}(\mu)=\sqrt{\varepsilon_{\rm B}(2\mu+\varepsilon_{\rm B})}\ \Theta(2\mu+\varepsilon_{\rm B}),
\end{equation}
which determines analytically the order parameter $\Delta$ as a function of the chemical potential $\mu$. Substituting this result into
$\Omega_{\rm MF}$, we obtain the mean-field grand canonical EOS
\begin{eqnarray}\label{MFGAP}
\Omega_{\rm MF}(\mu)=-\frac{m}{8\pi}(2\mu+\varepsilon_{\rm B})^2\Theta(2\mu+\varepsilon_{\rm B}).
\end{eqnarray}
The mean-field contribution to the particle density is given by
\begin{eqnarray}
n_{\rm MF}(\mu)=\frac{m}{2\pi}(2\mu+\varepsilon_{\rm B})\Theta(2\mu+\varepsilon_{\rm B}).
\end{eqnarray}
The above mean-field results show that the system undergoes a second-order quantum phase transition from the vacuum to a matter phase with nonvanishing density. The critical chemical potential is given by
\begin{equation}
\mu_c=-\frac{\varepsilon_{\rm B}}{2}.
\end{equation}

\subsection{Gaussian pair fluctuation theory}

Now let us include the quantum fluctuations. We include the Gaussian fluctuations only and approximate the effective action as
${\cal S}_{\text{eff}}[\Phi,\Phi^*]\simeq{\cal S}_{\rm MF}+{\cal S}_{\rm GF}[\phi,\phi^*]$.
The advantage of this Gaussian approximation is that the path integral over $\phi$ and $\phi^*$ can be carried out analytically. To evaluate the quadratic term ${\cal S}_{\rm GF}[\phi,\phi^*]$, we make the Fourier transformation
\begin{equation}
\phi(x)=\sqrt{\beta V}\sum_Q \phi(Q)e^{-iq_l\tau+i{\bf q}\cdot{\bf r}},
\end{equation}
where $Q=(iq_l,{\bf q})$, with $q_l=2l\pi  T$ ($l\in\mathbb{Z}$) being
the boson Matsubara frequency. We use the notation $\sum_Q=T\sum_{l}\sum_{\bf q}$ throughout. After some manipulations,
${\cal S}_{\rm GF}[\phi,\phi^*]$ can be expressed in a compact form,
\begin{eqnarray}
{\cal S}_{\rm GF}[\phi,\phi^*]=\frac{\beta V}{2}\sum_Q\left(\begin{array}{cc}
\phi^*(Q) & \phi(-Q)\end{array}\right){\bf M}(Q)\left(\begin{array}{cc} \phi(Q)\\
\phi^*(-Q)\end{array}\right).
\end{eqnarray}
The inverse boson propagator ${\bf M}(Q)$ is a $2\times 2$ matrix. At $T=0$, its elements are analytically given by
\begin{eqnarray}
&&{\bf M}_{11}(Q)={\bf M}_{22}(-Q)\nonumber\\
&=&\frac{1}{U}+\sum_{\bf k}\left(\frac{u_{\bf k}^2u_{{\bf k}+{\bf q}}^2}
{iq_l-E_{{\bf k}}-E_{{\bf k}+{\bf q}}}-\frac{\upsilon_{\bf k}^2\upsilon_{{\bf k}+{\bf q}}^2}
{iq_l+E_{{\bf k}}+E_{{\bf k}+{\bf q}}}\right),\nonumber\\
&&{\bf M}_{12}(Q)={\bf M}_{21}(Q)\nonumber\\
&=&\sum_{\bf k}\left(\frac{u_{\bf k}\upsilon_{\bf k}u_{{\bf k}+{\bf q}}\upsilon_{{\bf k}+{\bf q}}}
{iq_l+E_{{\bf k}}+E_{{\bf k}+{\bf q}}}-\frac{u_{\bf k}\upsilon_{\bf k}u_{{\bf k}+{\bf q}}\upsilon_{{\bf k}+{\bf q}}}
{iq_l-E_{{\bf k}}-E_{{\bf k}+{\bf q}}} \right).
\end{eqnarray}
Here the BCS distribution functions are defined as $\upsilon_{\bf k}^2=(1-\xi_{\bf k}/E_{\bf k})/2$ and $u_{\bf k}^2=1-\upsilon_{\bf k}^2$.
Note that Eq. (\ref{URG}) should be used to eliminate the UV divergence.

Considering the Gaussian fluctuations only, the partition function is approximated as
\begin{eqnarray}
{\cal Z}\simeq\exp{\left(-{\cal S}_{\rm MF}\right)}
\int[d\phi][d\phi^*]\exp\Big\{-{\cal S}_{\rm GF}[\phi,\phi^*]\Big\}.
\end{eqnarray}
Carrying out the path integral over $\phi$ and $\phi^*$, we obtain the grand potential $\Omega=\Omega_{\rm MF}+\Omega_{\rm GF}$,  where the contribution from the Gaussian fluctuations can be formally expressed as
\begin{eqnarray}
\Omega_{\rm GF}=\frac{1}{2}\sum_{Q}\ln\det{\bf M}(Q).
\end{eqnarray}
However, this formal expression is divergent because the convergent factors are not appropriately considered. Considering the convergent factors leads to a finite result~\cite{TH03,TH04}:
\begin{eqnarray}
\Omega_{\rm GF}=\frac{1}{2}\sum_{Q}\ln\left[\frac{{\bf
M}_{11}(Q)}{{\bf M}_{22}(Q)}\det{\bf M}(Q)\right]e^{iq_l0^+}.
\end{eqnarray}
The Matsubara frequency sum can be converted to a standard contour integral. At $T=0$, we have
\begin{eqnarray}
\Omega_{\rm GF}=\frac{1}{2}\sum_{{\bf q}}\int_{-\infty}^0\frac{d\omega}{\pi}\left[\delta_{\rm M}(\omega,{\bf q})
+\delta_{11}(\omega,{\bf q})-\delta_{22}(\omega,{\bf q})\right],
\end{eqnarray}
where the phase shifts are defined as $\delta_{\rm M}(\omega,{\bf q})=-{\rm Im}\ln\det {\bf M}(\omega+i\epsilon,{\bf q})$, $\delta_{11}(\omega,{\bf q})=-{\rm Im}\ln {\bf M}_{11}(\omega+i\epsilon,{\bf q})$, and $\delta_{22}(\omega,{\bf q})=-{\rm Im}\ln {\bf M}_{22}(\omega+i\epsilon,{\bf q})$.

A crucial element of the Gaussian pair fluctuation (GPF) theory is that the order parameter $\Delta$ should be determined by the extreme
of the mean-field grand potential $\Omega_{\rm MF}$ rather than the full grand potential $\Omega=\Omega_{\rm MF}+\Omega_{\rm GF}$~\cite{TH03,TH04}. Therefore, we still use the mean-field gap equation or the analytical result, (\ref{MFGAP}). The advantages of the use of the mean-field gap equation can be summarized as follows:
\\ (i) The mean-field solution for the order parameter, (\ref{MFGAP}), guarantees Goldstone's theorem. The dispersion of the Goldstone mode can be obtained by solving the equation
\begin{eqnarray}
\det {\bf M}(\omega,{\bf q})=0
\end{eqnarray}
for $\omega$ smaller than the two-particle continuum. The use of the mean-field solution, (\ref{MFGAP}), for the order parameter ensures that
$\det {\bf M}(0,{\bf 0})=0$. Therefore, the lightest collective mode is gapless and has a linear dispersion at low momentum ${\bf q}$.
We expect that the most important contribution from the quantum fluctuations is the Goldstone mode fluctuation. The use of the mean-field gap equation ensures that the Goldstone mode is gapless and hence enables us to take into account correctly the contribution from the Goldstone mode.
\\ (ii) The use of the mean-field solution, (\ref{MFGAP}), maintains the famous Silver Blaze property~\cite{Blave1,Blave2} even if we consider the contributions from the quantum fluctuations. Even though the critical chemical potential $\mu_c=-\varepsilon_{\rm B}/2$ for the vacuum-matter transition is obtained from the mean-field approximation, we expect that it is exact because the minimal chemical potential to create a bound state is exactly $2\mu_c=-\varepsilon_{\rm B}$. For $\mu<\mu_c$ and at $T=0$, the system stays in the vacuum phase with vanishing pressure and density. This is known as the Silver Blaze problem~~\cite{Blave1,Blave2}. Obviously, the mean-field EOS satisfies this property. Now we show that the Gaussian contribution $\Omega_{\rm GF}$ also satisfies this property. For $\mu<\mu_c$, we have $\Delta=0$ and hence $\Omega_{\rm GF}$ is given  by
\begin{eqnarray}
\Omega_{\rm GF}=\sum_{Q}\ln {\bf M}_0(iq_l,{\bf q})e^{iq_l0^+},
\end{eqnarray}
where the pair susceptibility ${\bf M}_0(iq_l,{\bf q})$ in the vacuum phase is analytically given by
\begin{eqnarray}\label{SUS}
{\bf M}_0(iq_l,{\bf q})&=&\sum_{\bf k}\left(\frac{1}
{iq_l-\xi_{{\bf k}+{\bf q}/2}-\xi_{{\bf k}-{\bf q}/2}}+\frac{1}{2\varepsilon_{\bf k}+\varepsilon_{\rm B}}\right)\nonumber\\
&=&\frac{m}{4\pi}\ln\left(\frac{-iq_l+\frac{{\bf q}^2}{4m}-2\mu}{\varepsilon_{\rm B}}\right).
\end{eqnarray}
At $T=0$, we obtain
\begin{eqnarray}
\Omega_{\rm GF}=\sum_{{\bf q}}\int_{-\infty}^0\frac{d\omega}{\pi}\delta_0(\omega,{\bf q}),
\end{eqnarray}
where $\delta_0(\omega,{\bf q})=-{\rm Im}\ln {\bf M}_0(\omega+i\epsilon,{\bf q})$. It is easy to show that $\delta_0(\omega,{\bf q})=0$ for $\omega<0$ in the vacuum phase $\mu<\mu_c$. Therefore, we have exactly $\Omega_{\rm GF}=0$ for $\mu<\mu_c$. Accordingly, the particle density
also vanishes in the vacuum.

\subsection{Imaginary frequency integration formalism}

For the Gaussian contribution $\Omega_{\rm GF}$, it is convenient to employ an alternative formalism which automatically satisfies the Silver Blaze property and also leads to faster convergence for numerical calculations. To this end, we define two functions, ${\bf M}_{11}^{\rm C}(z,{\bf q})$ and
${\bf M}_{22}^{\rm C}(z,{\bf q})$~\cite{TH04}, which are given by
\begin{eqnarray}
{\bf M}^{\rm C}_{11}(z,{\bf q})={\bf M}^{\rm C}_{22}(-z,{\bf q})
=\frac{1}{U}+\sum_{\bf k}\frac{u_{\bf k}^2u_{{\bf k}+{\bf q}}^2}{z-E_{{\bf k}}-E_{{\bf k}+{\bf q}}}.
\end{eqnarray}
Using the gap equation, (\ref{MFGAPE}), and the fact that $u_{\bf k}^2<1$, we can show that ${\bf M}_{11}^{\rm C}(z,{\bf q})$ has no singularities or zeros in the left half-plane (${\rm Re}z<0$). Therefore, the Matsubara sum $\sum_{q_l}\ln {\bf M}_{11}^{\rm C}(iq_l,{\bf q})$ vanishes at $T=0$ since
$\ln {\bf M}_{11}^{\rm C}(z,{\bf q})$ has no singularities in the left-half plane. Therefore, the Gaussian contribution at $T=0$ can be expressed as~\cite{TH04}
\begin{eqnarray}
\Omega_{\rm GF}=\frac{1}{2}\sum_Q
\ln\left[\frac{{\bf M}_{11}(iq_l,{\bf q}){\bf M}_{22}(iq_l,{\bf q})-{\bf M}_{12}^2(iq_l,{\bf q})}
{{\bf M}_{11}^{\rm C}(iq_l,{\bf q}){\bf M}_{22}^{\rm C}(iq_l,{\bf q})}\right].
\end{eqnarray}

At $T=0$, we replace the discrete Matsubara frequency sum with a continuous integral over an imaginary frequency; i.e.,
\begin{eqnarray}
T\sum_{l=-\infty}^\infty X(iq_l)\rightarrow\int_{-\infty}^\infty\frac{d\omega}{2\pi}X(i\omega).
\end{eqnarray}
After some manipulations, we obtain
\begin{widetext}
\begin{eqnarray}\label{OMEGA-GF}
\Omega_{\rm GF}(\mu)=\sum_{{\bf q}}\int_0^\infty\frac{d\omega}{2\pi}
\ln\left[1-2\Delta^4(\mu)\frac{A(\omega,{\bf q})C(\omega,{\bf q})+\omega^2B(\omega,{\bf q})D(\omega,{\bf q})
+2F^2(\omega,{\bf q})}{A^2(\omega,{\bf q})+\omega^2B^2(\omega,{\bf q})}+\Delta^8(\mu)\frac{C^2(\omega,{\bf q})+\omega^2D^2(\omega,{\bf q})}{A^2(\omega,{\bf q})+\omega^2B^2(\omega,{\bf q})}\right],
\end{eqnarray}
where we have used the fact that the integrand is real and even in $\omega$. The functions $A,B,C,D,$ and $F$ are defined as
\begin{eqnarray}\label{GPFfun}
&&A(\omega,{\bf q})=\sum_{\bf k} \left[\frac{1}{2\varepsilon_{\bf k}+\varepsilon_{\rm B}}
-\frac{1}{4}\left(\frac{1}{E_{{\bf k}+{\bf q}/2}}+\frac{1}{E_{{\bf k}-{\bf q}/2}}\right)
\frac{(E_{{\bf k}+{\bf q}/2}+\xi_{{\bf k}+{\bf q}/2})(E_{{\bf k}-{\bf q}/2}+\xi_{{\bf k}-{\bf q}/2})}
{(E_{{\bf k}+{\bf q}/2}+E_{{\bf k}-{\bf q}/2})^2+\omega^2}\right],\nonumber\\
&&B(\omega,{\bf q})=\sum_{\bf k}\frac{1}{4E_{{\bf k}+{\bf q}/2}E_{{\bf k}-{\bf q}/2}}
\frac{(E_{{\bf k}+{\bf q}/2}+\xi_{{\bf k}+{\bf q}/2})(E_{{\bf k}-{\bf q}/2}+\xi_{{\bf k}-{\bf q}/2})}
{(E_{{\bf k}+{\bf q}/2}+E_{{\bf k}-{\bf q}/2})^2+\omega^2},\nonumber\\
&&C(\omega,{\bf q})=\sum_{\bf k}\frac{1}{4}\left(\frac{1}{E_{{\bf k}+{\bf q}/2}}+\frac{1}{E_{{\bf k}-{\bf q}/2}}\right)
\frac{1}{(E_{{\bf k}+{\bf q}/2}+\xi_{{\bf k}+{\bf q}/2})(E_{{\bf k}-{\bf q}/2}+\xi_{{\bf k}-{\bf q}/2})}
\frac{1}{(E_{{\bf k}+{\bf q}/2}+E_{{\bf k}-{\bf q}/2})^2+\omega^2},\nonumber\\
&&D(\omega,{\bf q})=\sum_{\bf k} \frac{1}{4E_{{\bf k}+{\bf q}/2}E_{{\bf k}-{\bf q}/2}(E_{{\bf k}+{\bf q}/2}+\xi_{{\bf k}+{\bf q}/2})
(E_{{\bf k}-{\bf q}/2}+\xi_{{\bf k}-{\bf q}/2})}
\frac{1}{(E_{{\bf k}+{\bf q}/2}+E_{{\bf k}-{\bf q}/2})^2+\omega^2},\nonumber\\
&&F(\omega,{\bf q})=\sum_{\bf k}\frac{1}{4}\left(\frac{1}{E_{{\bf k}+{\bf q}/2}}+\frac{1}{E_{{\bf k}-{\bf q}/2}}\right)\frac{1}{(E_{{\bf k}+{\bf q}/2}+E_{{\bf k}-{\bf q}/2})^2+\omega^2}.
\end{eqnarray}
\end{widetext}
Note that $\Delta(\mu)$ is given by the mean-field solution, (\ref{MFGAP}). The BCS-type dispersions $E_{{\bf k}\pm{\bf q}/2}$ in (\ref{GPFfun}) are hence analytically given by
\begin{equation}
E_{{\bf k}\pm{\bf q}/2}=\sqrt{(\varepsilon_{{\bf k}\pm{\bf q}/2}-\mu)^2
+\varepsilon_{\rm B}(2\mu+\varepsilon_{\rm B})\Theta(2\mu+\varepsilon_{\rm B})}.
\end{equation}
The integrand in (\ref{OMEGA-GF}) vanishes in the vacuum $\mu<\mu_c$ and hence the Silver Blaze property is automatically satisfied. Moreover, because we use the mean-field gap equation, (\ref{MFGAP}), we can replace $1/(2\varepsilon_{\bf k}+\varepsilon_{\rm B})$ with $1/(2E_{\bf k})$ in the expression of the function $A(\omega,{\bf q})$. Then we find that the integrand in (\ref{OMEGA-GF}) diverges near $(\omega,{\bf q})=(0,{\bf 0})$, which indicates that the most important contribution is from the low-energy Goldstone mode.

In summary, the grand canonical EOS in the GPF theory is given by
\begin{equation}
\Omega(\mu)=\Omega_{\rm MF}(\mu)+\Omega_{\rm GF}(\mu).
\end{equation}
The particle density $n(\mu)$ reads
\begin{equation}
n(\mu)=n_{\rm MF}(\mu)+n_{\rm GF}(\mu),
\end{equation}
where the GF contribution is formally given by
\begin{equation}
n_{\rm GF}(\mu)=-\frac{d\Omega_{\rm GF}(\mu)}{d\mu}.
\end{equation}
Here $d/d\mu$ represents the full derivative with respect to $\mu$; i.e.,
\begin{equation}
\frac{d\Omega_{\rm GF}(\mu)}{d\mu}=\frac{\partial\Omega_{\rm GF}}{\partial\mu}+\frac{\partial\Omega_{\rm GF}}{\partial\Delta}\frac{d\Delta}{d\mu}.
\end{equation}
In 3D, it was shown that the second term is crucial to produce in the BEC limit the composite boson scattering length $a_{\rm B}=0.55a_{\rm 3D}$
~\cite{TH03,TH04}, which is very close to the result $a_{\rm B}=0.6a_{\rm 3D}$ from the exact four-body calculation~\cite{FB3D}. It has been shown that in 2D the second term is much more important than in 3D. Without this contribution, the fluctuation contribution to the particle density, $n_{\rm GF}(\mu)$, is divergent~\cite{2DTH15,N2D}. The full derivative leads to a convergent particle density and hence an appropriate description of the BCS-BEC crossover.

\section{Strong coupling limit: Weakly interacting 2D Bose Condensate}\label{s3}

While the mean-field theory predicts a simple and analytical EOS in the entire BCS-BEC crossover, it does not capture correctly the interaction between the composite bosons in the strong coupling (BEC) limit. At $\mu=\mu_c$, the system undergoes a second-order quantum phase transition from the vacuum to the dilute BEC of bound states. In the grand canonical ensemble, the BEC limit corresponds to the regime $\mu=\mu_c+0^+$, where the particle density $n$ is vanishingly small. Alternatively, the chemical potential for composite bosons is given by
\begin{equation}
\mu_{\rm B}=2\mu+\varepsilon_{\rm B}.
\end{equation}
The BEC limit corresponds to $\mu_{\rm B}\rightarrow0$ or, more explicitly, $\mu_{\rm B}/\varepsilon_{\rm B}\rightarrow0$.

First, we show that the mean-field theory leads to a constant coupling between the composite bosons~\cite{Coupling2D}. To this end, we derive the Gross-Pitaevskii free energy functional in the BEC limit~\cite{Blave2,GPFhe}. Since the order parameter becomes vanishingly small for $\mu=\mu_c+0^+$, we can obtain a Ginzburg-Landau free energy functional of the order parameter field $\Delta(x)$,
\begin{eqnarray}
\Omega_{\text{GL}}[\Delta]=\int dx\ \Bigg[\Delta^*\left(
a\frac{\partial}{\partial\tau}-b\frac{\mbox{\boldmath{$\nabla$}}^2}{4m}-c\right)\Delta+\frac{d}{2}|\Delta|^4\Bigg].
\end{eqnarray}
The coefficients $a,b,$ and $c$ can be determined by the normal-state pair susceptibility ${\bf M}_0(iq_l,{\bf q})$ which is given by (\ref{SUS}).
For $q_l,{\bf q}^2/(4m)\ll \varepsilon_{\rm B}$ and $\mu_{\rm B}\rightarrow0^+$, we have
\begin{eqnarray}
{\bf M}_0(iq_l,{\bf q})\simeq\frac{m}{4\pi\varepsilon_{\rm B}}\left(-iq_l+\frac{{\bf q}^2}{4m}-\mu_{\rm B}\right).
\end{eqnarray}
Therefore, we obtain
\begin{equation}
a=b=\frac{m}{4\pi\varepsilon_{\rm B}},\ \ \ \ c=\frac{m\mu_{\rm B}}{4\pi\varepsilon_{\rm B}}.
\end{equation}
The coefficient $d$ contains the information of the interaction between the composite bosons. In the mean-field theory, it can be obtained by making the Taylor expansion of $\Omega_{\rm MF}$ near $\Delta=0$. We obtain
\begin{equation}
d=\frac{m}{4\pi\varepsilon_{\rm B}^2}
\end{equation}
Therefore, if we define a new condensate wave function
\begin{eqnarray}
\varphi(x)=\sqrt{\frac{m}{4\pi\varepsilon_{\rm B}}}\Delta(x),
\end{eqnarray}
the Ginzgurg-Landau free energy reduces to the Gross-Pitaevskii free energy of a dilute Bose gas,
\begin{eqnarray}
\Omega_{\text{GP}}[\varphi]=\int dx\ \Bigg[\varphi^*\left(
\frac{\partial}{\partial\tau}-\frac{\mbox{\boldmath{$\nabla$}}^2}{2m_{\rm B}}
-\mu_{\rm B}\right)\varphi+\frac{g_{\rm B}}{2}|\varphi|^4\Bigg].
\end{eqnarray}
Here $m_{\rm B}=2m$ is the mass of the composite bosons. In the mean-field theory, the boson-boson coupling $g_{\rm B}$
is a constant,
\begin{eqnarray}\label{2Dcoupling}
g_{\rm B}=\frac{4\pi}{m}.
\end{eqnarray}
This result is consistent with the previous calculation above the superfluid transition temperature~\cite{Coupling2D}. However, it has been shown that for 2D bosons, the coupling $g_{\rm B}$ is energy (chemical potential) dependent~\cite{2DTH01,GPE2D}, that is,
\begin{equation}
\frac{1}{g_{\rm B}}=\frac{m_{\rm B}}{4\pi}\ln\left(\frac{4}{\mu_{\rm B}m_{\rm B}a_{\rm B}^2e^{2\gamma}}\right),
\end{equation}
where $a_{\rm B}$ is the boson-boson scattering length and $\gamma\simeq0.577...$ is the Euler constant. The constant coupling, (\ref{2Dcoupling}), indicates that the BEC limit of the 2D mean-field theory corresponds to the Born approximation for four-body scattering in 2D~\cite{FourB}. This is also true for 3D. However, in 3D, the Born approximation already predicts a weak coupling $g_{\rm B}=4\pi a_{\rm B}/m_{\rm B}$ with $a_{\rm B}=2a_{\rm 3D}$, and hence the 3D mean-field theory is qualitatively correct.

Second, the incorrect boson-boson interaction can also be seen from the EOS. In the mean-field theory, the grand canonical EOS in the BEC limit can be expressed as
\begin{eqnarray}
\Omega_{\rm MF}(\mu_{\rm B})=-\frac{m_{\rm B}}{16\pi}\mu_{\rm B}^2.
\end{eqnarray}
However, it is known that the grand canonical EOS of weakly interacting 2D Bose gases in the Bogoliubov theory is given by~\cite{2DBEC06}
\begin{eqnarray}\label{BEC2D}
\Omega(\mu_{\rm B})=-\frac{m_{\rm B}\mu_{\rm B}^2}{8\pi}\left[\ln\left(\frac{4}{\mu_{\rm B}m_{\rm B}a_{\rm B}^2e^{2\gamma+1}}\right)+\frac{1}{2}\right].
\end{eqnarray}
It was shown that the corrections beyond the Bogoliubov theory can be expanded in powers of the small parameter
$1/\ln[4/(\mu_{\rm B}m_{\rm B}a_{\rm B}^2e^{2\gamma+1})]$. The leading-order correction was presented in~\cite{2DBEC06}. In the BEC limit $\mu_{\rm B}\rightarrow0$, the beyond-Bogoliubov corrections are vanishingly small in comparison to the Bogoliubov contribution. We expect that the Bogoliubov EOS, (\ref{BEC2D}), can be recovered in the BEC limit $\mu_{\rm B}/\varepsilon_{\rm B}\rightarrow0$ if we include the contribution from the Gaussian quantum fluctuations. If so, this allows us to determine the composite boson scattering length by comparing our EOS with the Bogoliubov EOS, (\ref{BEC2D}), in the limit $\mu_{\rm B}/\varepsilon_{\rm B}\rightarrow0$.

In the GPF theory, the grand canonical EOS is given by
\begin{equation}\label{EOS-BEC}
\Omega(\mu_{\rm B})=-\frac{m_{\rm B}\mu_{\rm B}^2}{8\pi}\left[f(\zeta)+\frac{1}{2}\right],
\end{equation}
where $\zeta=\mu_{\rm B}/\varepsilon_{\rm B}$ and the function $f(\zeta)$ is given by
\begin{widetext}
\begin{eqnarray}\label{FZETA}
f(\zeta)=-\frac{2}{\pi\zeta^2}\int_0^\infty dx\int_0^\infty dy
\ln\left[1-2\zeta^2\frac{{\cal A}(x,y){\cal C}(x,y)+y^2{\cal B}(x,y){\cal D}(x,y)+2{\cal F}^2(x,y)}{{\cal A}^2(x,y)+y^2{\cal B}^2(x,y)}
+\zeta^4\frac{{\cal C}^2(x,y)+y^2{\cal D}^2(x,y)}{{\cal A}^2(x,y)+y^2{\cal B}^2(x,y)}\right].
\end{eqnarray}
Here the dimensionless functions ${\cal A},{\cal B},{\cal C},{\cal D},$ and ${\cal F}$ are given by
\begin{eqnarray}
&&{\cal A}(x,y)=\int_0^{2\pi}\frac{d\theta}{2\pi}\int_0^\infty dz\ \left[\frac{1}{z+1}-\frac{1}{4}\left(\frac{1}{E_+}+\frac{1}{E_-}\right)\frac{(E_++\xi_+)(E_-+\xi_-)}{(E_++E_-)^2+y^2}\right],\nonumber\\
&&{\cal B}(x,y)=\int_0^{2\pi}\frac{d\theta}{2\pi}\int_0^\infty dz\ \frac{1}{4E_+E_-}\frac{(E_++\xi_+)(E_-+\xi_-)}{(E_++E_-)^2+y^2},\nonumber\\
&&{\cal C}(x,y)=\int_0^{2\pi}\frac{d\theta}{2\pi}\int_0^\infty dz\ \frac{1}{4}\left(\frac{1}{E_+}+\frac{1}{E_-}\right)
\frac{1}{(E_++\xi_+)(E_-+\xi_-)}\frac{1}{(E_++E_-)^2+y^2},\nonumber\\
&&{\cal D}(x,y)=\int_0^{2\pi}\frac{d\theta}{2\pi}\int_0^\infty dz\ \frac{1}{4E_+E_-(E_++\xi_+)(E_-+\xi_-)}\frac{1}{(E_++E_-)^2+y^2},\nonumber\\
&&{\cal F}(x,y)=\int_0^{2\pi}\frac{d\theta}{2\pi}\int_0^\infty dz\ \frac{1}{4}\left(\frac{1}{E_+}+\frac{1}{E_-}\right)\frac{1}{(E_++E_-)^2+y^2},
\end{eqnarray}
where the dimensionless variables $x,y,$ and $z$ are defined as $x={\bf q}^2/(4m\varepsilon_{\rm B})$, $y=\omega/\varepsilon_{\rm B}$, and
$z={\bf k}^2/(m\varepsilon_{\rm B})$ and the notations $\xi_\pm$ and $E_\pm$ are given by
\begin{eqnarray}
\xi_\pm=\frac{1}{2}\left(z+x\pm2\sqrt{xz}\cos\theta+1-\zeta\right),\ \ \ \ \ \ \ \
E_\pm=\sqrt{(\xi_\pm)^2+\zeta}.
\end{eqnarray}
\end{widetext}
We can show that $f(\zeta)$ is divergent at $\zeta=0$. To this end, we evaluate the functions
${\cal A},{\cal B},{\cal C},{\cal D},$ and ${\cal F}$ at $\zeta=0$, which is denoted by the subscript $0$. We have
\begin{eqnarray}
&&{\cal A}_0(x,y)=\frac{1}{2}\ln\left[(1+x)^2+y^2\right],\nonumber\\
&&{\cal B}_0(x,y)=\frac{1}{y}\arctan\frac{y}{1+x},\nonumber\\
&&{\cal C}_0(x,y)=\int_0^\infty dz\frac{L^2-2xz}{L^2(L^2+y^2)\left(L^2-4xz\right)^{3/2}},\nonumber\\
&&{\cal D}_0(x,y)=\int_0^\infty dz\frac{L^2-2xz}{L^3(L^2+y^2)\left(L^2-4xz\right)^{3/2}},\nonumber\\
&&{\cal F}_0(x,y)=\int_0^\infty dz\frac{1}{(L^2+y^2)\sqrt{L^2-4xz}}.
\end{eqnarray}
Here we define $L\equiv z+1+x$ for convenience. In the infrared limit, $x\rightarrow0$ and $y\rightarrow0$, the above functions behave as
\begin{eqnarray}\label{IR}
&&{\cal A}_0(x,y)\simeq x,\ \ \ \ \ \ {\cal B}_0(x,y)\simeq1,\nonumber\\
&&{\cal C}_0(x,y)\simeq\frac{1}{4},\ \ \ \ \ \ {\cal D}_0(x,y)\simeq\frac{1}{5},\nonumber\\
&&{\cal F}_0(x,y)\simeq\frac{1}{2}.
\end{eqnarray}
For further analysis it is convenient to employ the polar coordinates $x=\rho\cos\varphi$ and $y=\rho\sin\varphi$. By making use of the Taylor expansion for the logarithm in (\ref{FZETA}) (see Appendix A), we find that at precisely $\zeta=0$, the function $f(\zeta)$ is divergent because of the infrared behavior ${\cal A}_0^2+y^2{\cal B}_0^2\simeq \rho^2$. We note that this kind of divergence does not exist in 3D. In 3D, the mean-field theory already predicts a weakly interacting Bose condensate in the strong coupling limit with a composite boson scattering length
$a_{\rm B}=2a_{\rm 3D}$~\cite{BCSBEC2}. The inclusion of the Gaussian contribution in the BEC limit leads to a modification of the composite boson scattering length from the mean-field value $2a_{\rm 3D}$ to $0.55a_{\rm 3D}$~\cite{TH03,TH04}.

The divergence of the function $f(\zeta)$ at $\zeta\rightarrow0$ is not surprising. It is actually consistent with the Bogoliubov EOS (\ref{BEC2D}) where the logarithmic term in the brackets diverges when $\mu_{\rm B}\rightarrow0$. Therefore, we expect that for $\zeta\rightarrow0$, the function $f(\zeta)$ diverges as $-\ln\zeta=\ln(\varepsilon_{\rm B}/\mu_{\rm B})$. To show this logarithmic divergence, we separate the function $f(\zeta)$ into a divergent piece and a finite piece. The details are presented in Appendix A. The divergent piece is given by
\begin{eqnarray}\label{FZETA-APP}
f_{\rm d}(\zeta)=\frac{8}{\pi}\int_0^\infty dx\int_0^\infty dy
\frac{{\cal F}^2}{{\cal A}^2+y^2{\cal B}^2}.
\end{eqnarray}
To capture the asymptotic behavior of this divergent piece for $\zeta\rightarrow0$, we find that it is sufficient to expand
the denominator ${\cal A}^2+y^2{\cal B}^2$ to the order $O(\zeta^2)$ and approximate it as
\begin{equation}
{\cal A}^2+y^2{\cal B}^2\simeq {\cal J}(x,y)={\cal A}_0^2+y^2{\cal B}_0^2+2\zeta{\cal A}_0{\cal A}_1+\zeta^2{\cal A}_1^2.
\end{equation}
The explicit form of the function ${\cal A}_1(x,y)$ is shown in Appendix A. In the infrared limit $\rho\rightarrow0$, we have ${\cal A}_1\simeq1$.
The neglected terms in the above $\zeta$ expansion lead to vanishing contributions for $\zeta\rightarrow0$. Therefore, the infrared divergence in the limit $\zeta\rightarrow0$ behaves as
\begin{eqnarray}
\frac{8}{\pi}\int_0^{\pi/2}d\varphi\int_0^\epsilon\rho d\rho
\frac{1/4}{\rho^2+2\zeta\rho\cos\varphi+\zeta^2}\sim\ln\frac{\varepsilon_{\rm B}}{\mu_{\rm B}}.
\end{eqnarray}
Thus we have shown that in the BEC limit $\mu_{\rm B}\rightarrow0$, the Gaussian contribution $\Omega_{\rm GF}$ behaves exactly like the logarithmic term in the Bogoliubov EOS, (\ref{BEC2D}).

To obtain the composite boson scattering length $a_{\rm B}$, we need to determine the finite piece $\lambda$, which can be defined as
\begin{equation}\label{LAMBDA}
\lambda=\lim_{\zeta\rightarrow0}\left[f(\zeta)+\ln\zeta\right].
\end{equation}
Using the definition of the fermion scattering length $a_{\rm 2D}$,
\begin{equation}
\varepsilon_{\rm B}=\frac{4}{ma_{\rm 2D}^2e^{2\gamma}},
\end{equation}
we obtain the composite boson scattering length
\begin{equation}
a_{\rm B}=\kappa a_{\rm 2D},\ \ \ \ \kappa=\sqrt{\frac{1}{2e^{1+\lambda}}}.
\end{equation}
A careful numerical analysis (see Appendix A) shows that $\lambda\simeq-0.54$. Therefore, we obtain
\begin{equation}
\kappa\simeq0.56.
\end{equation}
This result is in good agreement with $\kappa\simeq0.56$ from the exact four-body calculation~\cite{2DTH01} and $\kappa\simeq0.55(4)$ from the EOS predicted by the diffusion Monte Carlo simulation~\cite{QMC2D1}. We also notice that the pole approximation of the Gaussian quantum fluctuations
with a dimensional regularization of the UV divergence in the BEC limit predicted an analytical result $\lambda=-1/2$ and hence $\kappa=1/(2^{1/2}e^{1/4})\simeq0.55$~\cite{2DTH24}.

\section{BCS-BEC crossover}\label{s4}

In this section, we study numerically the EOS in the entire BCS-BEC crossover. The determination of the grand canonical EOS is simple. The grand potential $\Omega(\mu)$ can be obtained by performing the numerical integration in (\ref{OMEGA-GF}) for $-\varepsilon_{\rm B}/2<\mu<+\infty$. The BEC and BCS limits corresponds to $\mu\rightarrow-\varepsilon_{\rm B}/2$ and $\mu\rightarrow+\infty$, respectively. In this work, we are interested in the canonical EOS for a homogeneous 2D Fermi gas with fixed density $n$. This enables us to compare our results with recent quantum Monte Carlo calculations of the energy density~\cite{QMC2D1,QMC2D2} and experimental measurements of the local pressure~\cite{2Dexp2,2Dexp6}. For convenience, we  define the Fermi momentum $k_{\text F}$ and the Fermi energy $\varepsilon_{\rm F}$ for a noninteracting 2D Fermi gas with the same density $n$. They are given by $k_{\rm F}=\sqrt{2\pi n}$ and $\varepsilon_{\rm F}=\pi n/m$. The BCS-BEC crossover is controlled by the dimensionless ratio $\alpha=\varepsilon_{\rm B}/\varepsilon_{\rm F}$ or the gas parameter
\begin{equation}
\eta=\ln(k_{\rm F}a_{\rm 2D}).
\end{equation}
The BCS and BEC limits correspond to $\eta\rightarrow+\infty$ and $\eta\rightarrow-\infty$, respectively.

In the mean-field approximation, we have $n\simeq n_{\rm MF}(\mu)$, which gives rise to the mean-field results of the chemical potential and the pairing gap~\cite{BCSBEC2D-2,BCSBEC2D-3},
\begin{equation}
\mu_{\rm MF}(n)=\varepsilon_{\rm F}-\frac{\varepsilon_{\rm B}}{2},\ \ \ \ \ \ \ \Delta_{\rm MF}(n)=\sqrt{2\varepsilon_{\rm B}\varepsilon_{\rm F}}.
\end{equation}
The energy density and pressure in the mean-field theory are given by
\begin{eqnarray}
&&E_{\rm MF}(n)=\Omega_{\rm MF}(\mu_{\rm MF})+\mu_{\rm MF} n=E_{\rm FG}-\frac{1}{2}n\varepsilon_{\rm B},\nonumber\\
&&P_{\rm MF}(n)=-\Omega_{\rm MF}(\mu_{\rm MF})=P_{\rm FG},
\end{eqnarray}
where $E_{\rm FG}=n\varepsilon_{\rm F}/2$ and $P_{\rm FG}=n\varepsilon_{\rm F}/2$ are the energy density and pressure of a noninteracting 2D Fermi gas with density $n$, respectively. We see clearly from the pressure that the mean-field theory does not recover a weakly interacting Bose condensate in the strong attraction limit.

To show that the chemical potential and the energy density suffer from the same problem, we define two dimensionless quantities
\begin{equation}
\nu=\frac{\mu+\varepsilon_{\rm B}/2}{\varepsilon_{\rm F}},\ \ \ \ \ \ \ \ \
R=\frac{E+n\varepsilon_{\rm B}/2}{E_{\rm FG}}.
\end{equation}
In the mean-field theory, the solutions of $\nu$ and $R$ are independent of the attraction strength in the entire BCS-BEC crossover; i.e.,
\begin{equation}\label{RMF}
\nu_{\rm MF}=1,\ \ \ \ \ \ \ \ \ R_{\rm MF}=1.
\end{equation}
On the other hand, the Bogoliubov theory predicts that the canonical EOS of a 2D Bose gas is given by
~\cite{2DBEC01,2DBEC02,2DBEC03,2DBEC04,2DBEC05,2DBEC06,2DBEC07}
\begin{eqnarray}\label{EOSBC}
\mu_{\rm B}=\frac{4\pi n_{\rm B}}{m_{\rm B}}\frac{1}{\ln\left(\frac{1}{n_{\rm B}a_{\rm B}^2}\right)},\ \ \ \ \
E=-n_{\rm B}\varepsilon_{\rm B}+\frac{2\pi n_{\rm B}^2}{m_{\rm B}}\frac{1}{\ln\left(\frac{1}{n_{\rm B}a_{\rm B}^2}\right)},
\end{eqnarray}
where $n_{\rm B}=n/2$ is the density of tightly bound bosons. Therefore, we expect that in the BEC limit ($\eta\rightarrow-\infty$) the solutions of $\nu$ and $R$ behave asymptotically as
\begin{eqnarray}
\nu\sim\frac{1}{2}\frac{1}{\ln\left(\frac{4\pi}{\kappa^2}\right)-2\eta},\ \ \ \ \ \ \ \ \ R\sim\frac{1}{2}\frac{1}{\ln\left(\frac{4\pi}{\kappa^2}\right)-2\eta},
\end{eqnarray}
where $\kappa\simeq0.56$ from the exact four-body calculation~\cite{2DTH01} or from our study in Sec. \ref{s3}. These results indicate that $\nu$ and $R$ become vanishingly small in the BEC limit. We note that the use of the Bogoliubov EOS (\ref{EOSBC}) requires that the parameter $1/\ln[1/(n_{\rm B}a_{\rm B}^2)]$ is sufficiently small or $\eta\rightarrow-\infty$. The corrections beyond the Bogoliubov theory was
studied in Refs.~\cite{2DBEC01,2DBEC02,2DBEC03,2DBEC04,2DBEC05,2DBEC06,2DBEC07}. On the other hand, in the BCS limit ($\eta\rightarrow+\infty$),
the pairing gap $\Delta$ becomes vanishingly small and hence the GPF theory becomes equivalent to the particle-particle ladder resummation~\cite{TH01,TH02,TH03,TH04}. Therefore, in the BCS limit, the GPF theory naturally recovers the perturbative EOS of a weakly interacting 2D Fermi gas up to the order $O(1/\eta^2)$. The perturbative EOS of a weakly interacting 2D Fermi gas is given by~\cite{PEOS2D1,PEOS2D2,PEOS2D3}
\begin{eqnarray}\label{PEOSFL}
\nu&=&1-\frac{1}{\eta}+\frac{\gamma+1-2\ln2}{\eta^2}+O\left(\frac{1}{\eta^3}\right),\nonumber\\
R&=&1-\frac{1}{\eta}+\frac{\gamma+3/4-2\ln2}{\eta^2}+O\left(\frac{1}{\eta^3}\right).
\end{eqnarray}
Therefore, we expect that $\nu$ and $R$ approach unity asymptotically for $\eta\rightarrow+\infty$.

In the GPF theory, the chemical potential $\mu$ is determined by solving the full number equation
\begin{equation}
n=n_{\rm MF}(\mu)+n_{\rm GF}(\mu).
\end{equation}
Then we can determine the energy density $E(n)=\Omega_{\rm MF}(\mu)+\Omega_{\rm GF}(\mu)+\mu n$ and the pressure $P(n)=-\Omega_{\rm MF}(\mu)
-\Omega_{\rm GF}(\mu)$. The Gaussian contribution $n_{\rm GF}(\mu)$ can be worked out analytically but it is rather tedious. In practice, we start from the grand potential $\Omega(\mu)=\Omega_{\rm MF}(\mu)+\Omega_{\rm GF}(\mu)$. To determine the chemical potential $\mu$, we calculate the energy density as a function of $\mu$; i.e., $E(\mu)=\Omega(\mu)+\mu n$. We search for the maximum of $E(\mu)$, which gives rise to the solution of the chemical potential for the given density $n$. Meanwhile, the energy density and the pressure for the given density $n$ are determined. To perform the numerical calculation, it is convenient to use the dimensionless variable $\nu$. The mean-field contribution to the grand potential is $\Omega_{\rm MF}(\mu)=-\nu^2 E_{\rm FG}$. The Gaussian contribution to the grand potential can be expressed as
\begin{equation}
\Omega_{\rm g}(\mu)=g(\nu)E_{\rm FG},
\end{equation}
where the function $g(\nu)$ is given by
\begin{widetext}
\begin{eqnarray}
g(\nu)=\frac{2}{\pi}\int_0^\infty ds\int_0^\infty dt
\ln\left[1-8\alpha^2\nu^2\frac{{\cal A}(s,t){\cal C}(s,t)+t^2{\cal B}(s,t){\cal D}(s,t)+2{\cal F}^2(s,t)}{{\cal A}^2(s,t)+t^2{\cal B}^2(s,t)}
+16\alpha^4\nu^4\frac{{\cal C}^2(s,t)+t^2{\cal D}^2(s,t)}{{\cal A}^2(s,t)+t^2{\cal B}^2(s,t)}\right].
\end{eqnarray}
The dimensionless functions ${\cal A},{\cal B},{\cal C},{\cal D},$ and ${\cal F}$ are now defined as
\begin{eqnarray}
&&{\cal A}(s,t)=\int_0^{2\pi}\frac{d\theta}{2\pi}\int_0^\infty du\ \left[\frac{1}{2u+\alpha}-\frac{1}{4}\left(\frac{1}{E_+}+\frac{1}{E_-}\right)\frac{(E_++\xi_+)(E_-+\xi_-)}{(E_++E_-)^2+t^2}\right],\nonumber\\
&&{\cal B}(s,t)=\int_0^{2\pi}\frac{d\theta}{2\pi}\int_0^\infty du\ \frac{1}{4E_+E_-}\frac{(E_++\xi_+)(E_-+\xi_-)}{(E_++E_-)^2+t^2},\nonumber\\
&&{\cal C}(s,t)=\int_0^{2\pi}\frac{d\theta}{2\pi}\int_0^\infty du\ \frac{1}{4}\left(\frac{1}{E_+}+\frac{1}{E_-}\right)
\frac{1}{(E_++\xi_+)(E_-+\xi_-)}\frac{1}{(E_++E_-)^2+t^2},\nonumber\\
&&{\cal D}(s,t)=\int_0^{2\pi}\frac{d\theta}{2\pi}\int_0^\infty du\ \frac{1}{4E_+E_-(E_++\xi_+)(E_-+\xi_-)}\frac{1}{(E_++E_-)^2+t^2},\nonumber\\
&&{\cal F}(s,t)=\int_0^{2\pi}\frac{d\theta}{2\pi}\int_0^\infty du\ \frac{1}{4}\left(\frac{1}{E_+}+\frac{1}{E_-}\right)\frac{1}{(E_++E_-)^2+t^2},
\end{eqnarray}
\end{widetext}
where the variables $s={\bf q}^2/(8m\varepsilon_{\rm F})$, $t=\omega/\varepsilon_{\rm F}$, and $u={\bf k}^2/(2m\varepsilon_{\rm F})$. Here the notations $\xi_\pm$ and $E_\pm$ are given by
\begin{eqnarray}
\xi_\pm&=&u+s\pm2\sqrt{us}\cos\theta-\nu+\frac{\alpha}{2},\nonumber\\
E_\pm&=&\sqrt{(\xi_\pm)^2+2\alpha\nu}.
\end{eqnarray}

\begin{figure}[!htb]
\begin{center}
\includegraphics[width=9cm]{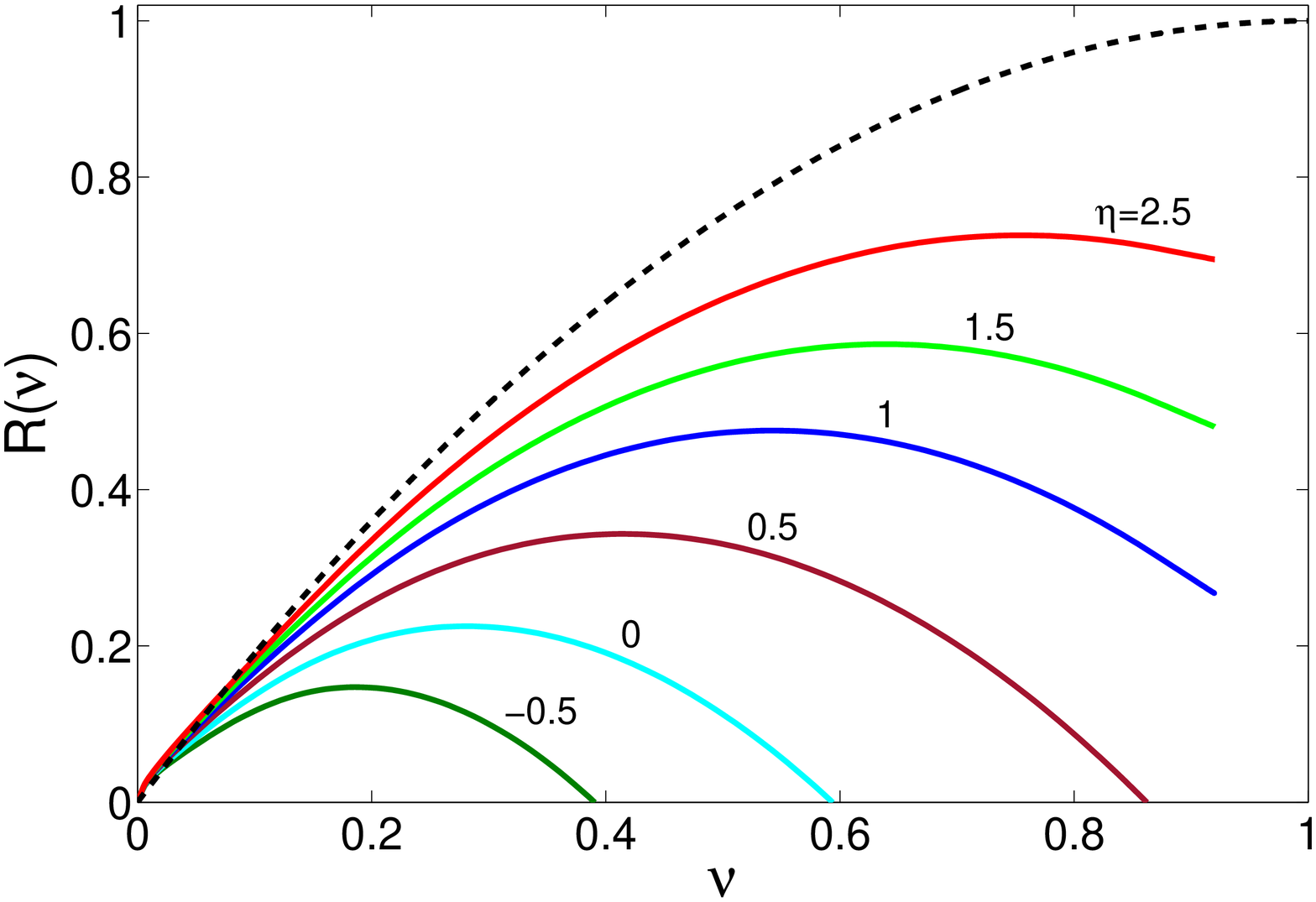}
\caption{(Color online) Curves of the function $R(\nu)$ for various values of the gas parameter $\eta=\ln(k_{\rm F}a_{\rm 2D})$.
For comparison, we show the mean-field prediction $R_{\rm MF}(\nu)=-\nu^2+2\nu$ by the dashed line. \label{fig1}}
\end{center}
\end{figure}

Using the function $g(\nu)$ we have defined, we can express the dimensionless quantity $R$ as
\begin{equation}
R(\nu)\equiv\frac{E(\mu)+\frac{1}{2}n\varepsilon_{\rm B}}{E_{\rm FG}}=-\nu^2+g(\nu)+2\nu.
\end{equation}
The physical results of $\nu$ and $R$ correspond to the maximum point of the the function $R(\nu)$ in the range $0\leq\nu\leq1$. In the mean-field theory, we neglect the Gaussian contribution $g(\nu)$ and hence $R(\nu)\simeq-\nu^2+2\nu$. The maximum of the function $R(\nu)$ gives the results
$\nu=1$ and $R=1$, which are precisely the mean-field predictions, (\ref{RMF}). Including the Gaussian contribution $g(\nu)$, the maximum of $R(\nu)$ will be modified since the function $g(\nu)$ depends explicitly on the interaction strength $\alpha$ or the gas parameter $\eta$. In Fig. \ref{fig1}, we show the curves of the function $R(\nu)$ for several values of the gas parameter $\eta$. We find that the quantum fluctuations become more and more important when the attraction strength increases.

\begin{figure}[!htb]
\begin{center}
\includegraphics[width=9cm]{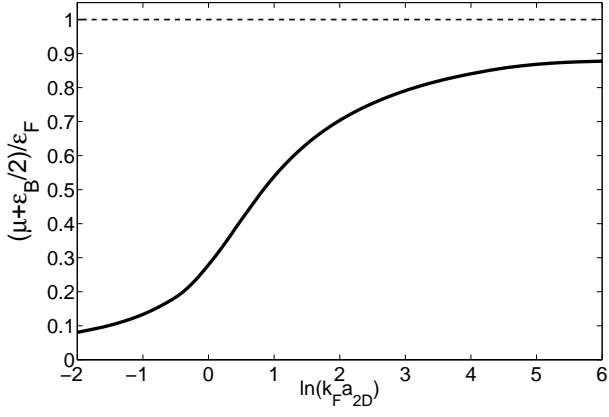}
\caption{Evolution of the chemical potential $\mu$ in the BCS-BEC crossover. We show the quantity
$\nu=(\mu+\varepsilon_{\rm B}/2)/\varepsilon_{\rm F}$ as a function of the gas parameter $\eta=\ln(k_{\rm F}a_{\rm 2D})$.
The mean-field prediction is represented by the dashed line. \label{fig2}}
\end{center}
\end{figure}

\begin{figure}[!htb]
\begin{center}
\includegraphics[width=9cm]{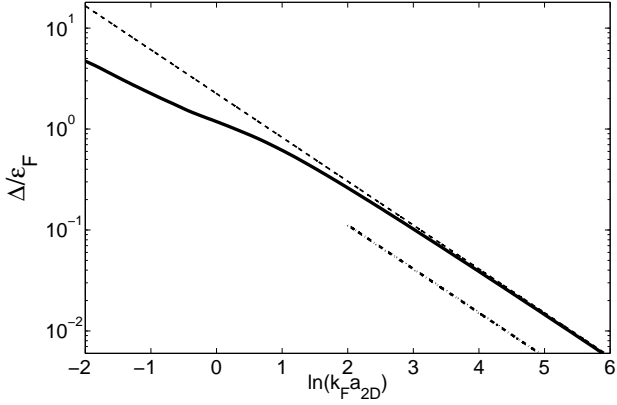}
\caption{The order parameter or pairing gap $\Delta$ (divided by $\varepsilon_{\rm F}$) as a function of the gas parameter
$\eta=\ln(k_{\rm F}a_{\rm 2D})$. The dashed line is the mean-field prediction. The dashed-dotted line corresponds to the
prediction with GMB effect in the weak coupling regime ($\eta>2$).\label{fig3}}
\end{center}
\end{figure}

\begin{figure}[!htb]
\begin{center}
\includegraphics[width=9cm]{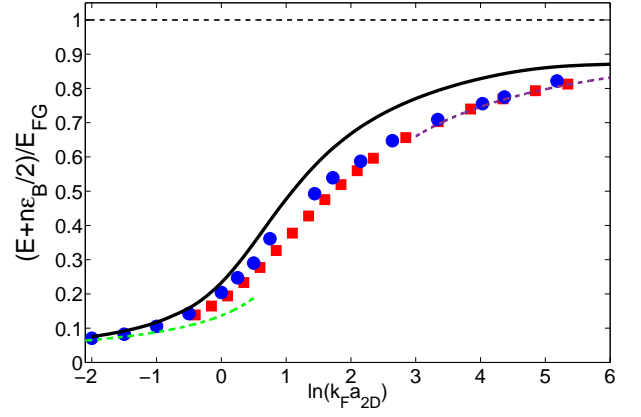}
\caption{(Color online) Evolution of the energy density $E$ in the BCS-BEC crossover. The quantity
$R=(E+n\varepsilon_{\rm B}/2)/E_{\rm FG}$ is shown as a function of the gas parameter $\eta=\ln(k_{\rm F}a_{\rm 2D})$.
The dashed line represents the mean-field prediction. The (blue) circles and (red) squares represent the predictions from the
diffusion Monte Carlo simulation~\cite{QMC2D1} and the auxiliary-field Monte Carlo simulation~\cite{QMC2D2}, respectively.
The bottom-left dashed (green) line represents the Bogoliubov EOS of a weakly interacting 2D Bose gas with the boson scattering length
$a_{\rm B}=0.56a_{\rm 2D}$ [see Eq. (\ref{EOSBC})]. The top-right dashed (purple) line shows the EOS of a weakly interacting 2D
Fermi gas [see Eq. (\ref{PEOSFL})].  \label{fig4}}
\end{center}
\end{figure}

\begin{figure}[!htb]
\begin{center}
\includegraphics[width=9cm]{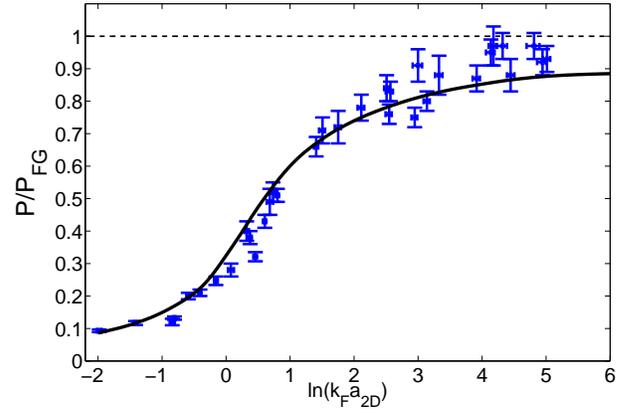}
\caption{(Color online) Evolution of the pressure $P$ in the BCS-BEC crossover. $P/P_{\rm FG}$ as a function of
the gas parameter $\eta=\ln(k_{\rm F}a_{\rm 2D})$ is shown. The mean-field prediction is represented by the dashed line.
The (blue) circles with error bars are the experimental data taken from \cite{2Dexp2}. \label{fig5}}
\end{center}
\end{figure}

\begin{figure}[!htb]
\begin{center}
\includegraphics[width=9cm]{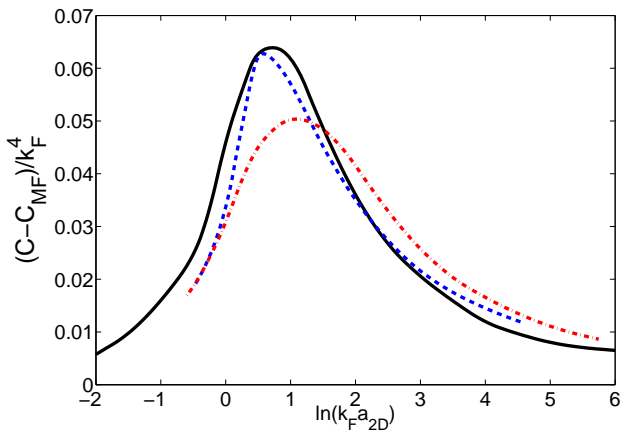}
\caption{(Color online) Evolution of the contact $C$ in the BCS-BEC crossover. This figure shows $(C-C_{\rm MF})/k_{\rm F}^4$ as a function of
the gas parameter $\eta=\ln(k_{\rm F}a_{\rm 2D})$, where $C_{\rm MF}=\alpha/2$ is the mean-field prediction. The dashed (blue) line
and the dash-dotted (red) line represent the predictions from the diffusion Monte Carlo simulation~\cite{QMC2D1} and the auxiliary-field 
Monte Carlo simulation~\cite{QMC2D2}, respectively.\label{fig6}}
\end{center}
\end{figure}

In Fig. \ref{fig2}, we show the evolution of the chemical potential $\mu$ or, explicitly, the quantity $\nu=(\mu+\varepsilon_{\rm B}/2)/\varepsilon_{\rm F}$ in the BCS-BEC crossover. We find that $\nu\rightarrow1$ in the BCS limit and $\nu\rightarrow0$ in the BEC limit, in agreement with our general expectation. The order parameter $\Delta$ is shown in Fig. \ref{fig3}. We find that the inclusion of the quantum fluctuations leads to a large suppression of the order parameter in the strong coupling regime. At weak coupling, it was shown that the induced interaction or the Gor'kov--Melik-Barkhudarov (GMB) effect~\cite{GMB} leads to a suppression of the critical temperature and hence the pairing gap $\Delta$ by a factor of $1/e$~\cite{2DTH01}. In Fig. \ref{fig3}, we also show the prediction with the GMB effect in the weak coupling regime ($\eta>2$). Obviously, the current GPF theory does not take into account the GMB effect.

In Fig. \ref{fig4}, we show the evolution of the energy density $E$ or, explicitly, the quantity $R=(E+n\varepsilon_{\rm B}/2)/E_{\rm FG}$ in the BCS-BEC crossover. In the BEC limit ($\eta\rightarrow-\infty$), our result approaches the Bogoliubov EOS, (\ref{EOSBC}), of weakly interacting 2D Bose gases with the boson scattering length $a_{\rm B}\simeq0.56a_{\rm 2D}$. In the BCS limit ($\eta\rightarrow+\infty$), our result tends to the perturbative EOS, (\ref{PEOSFL}), of weakly interacting 2D Fermi gases. The energy density was computed recently by using the diffusion Monte Carlo simulation~\cite{QMC2D1} and the auxiliary-field Monte Carlo simulation~\cite{QMC2D2}. In Fig. \ref{fig4}, we also show these Monte Carlo results for comparison. Even though our theory recovers the correct BCS and BEC limits, there exists a slight deviation between our theoretical prediction and the Monte Carlo results. This is not surprising because the GPF theory, which considers only the Gaussian pair fluctuations, is not an exact treatment. Some many-body effects we have not taken into account in the GPF theory may account for this disagreement. First, the current GPF theory does not consider the GMB effect~\cite{GMB}, which leads to a suppression of the pairing gap $\Delta$ by a factor of $1/e$ at weak coupling~\cite{2DTH01}. The inclusion of this effect may lead to a slight suppression of the energy density and a faster convergence to the EOS, (\ref{PEOSFL}), of weakly interacting 2D Fermi gases. The GMB effect may also be important in the crossover regime (roughly $-0.5<\eta<2$). Second, in the GPF theory, we consider only the Gaussian pair fluctuations. The contributions from the non-Gaussian quantum fluctuations (beyond quadratic order in $\phi$ and $\phi^*$) may be important to make for a better agreement with the Monte Carlo results in the crossover regime.

As we have mentioned, the most important thermodynamic quantity which shows the significance of the quantum fluctuations is the pressure $P$. The mean-field theory predicts $P=P_{\rm GF}$ for arbitrary attraction strength. In the GPF theory, we have
\begin{equation}
\frac{P(n)}{P_{\rm FG}}=2\nu-R,
\end{equation}
where $\nu$ and $R$ have been determined by searching for the maximum of the function $R(\nu)$. Therefore, the pressure depends explicitly on the interaction strength. In Fig. \ref{fig5}, we show the evolution of the pressure or, explicitly, the ratio $P/P_{\rm FG}$ in the BCS-BEC crossover. Recent experiments on the quasi-2D Fermi gases across a Feshbach resonance have measured the local pressure at the center of the atom trap at sufficiently low temperatures~\cite{2Dexp2,2Dexp6}, which can be regarded as the ground-state pressure of a homogeneous 2D Fermi gas in the BCS-BEC crossover. In Fig. \ref{fig5}, we also show the experimental data reported in~\cite{2Dexp2}. Except for the deep BCS regime ($\eta>3$), our theoretical prediction is in good agreement with the experimental measurement. The observed high pressure in the deep BCS regime could be attributed
to the mesoscopic nature of the experimental system: In the deep BCS regime, the scattering length $a_{\rm 2D}$ becomes larger than the cloud size and hence the interaction is effectively suppressed~\cite{2Dexp2}. On the other hand, it has been argued that the temperature effect may also be crucial to understand the observed high pressure in the deep BCS regime~\cite{2DTH21}. In the future, it is necessary to study the finite-temperature effect in the current GPF theory.

Having determined the EOS, we can calculate the contact $C$, which is a powerful quantity to relate the energy, pressure, and the microscopic momentum distribution~\cite{Contact01,Contact02}. In 2D, the contact $C$ can be defined as~\cite{Contact2D}
\begin{equation}
\frac{C}{k_{\rm F}^4}=\frac{1}{4}\frac{d(E/E_{\rm FG})}{d\eta}.
\end{equation}
After some simple manipulation, we obtain
\begin{equation}
\frac{C}{k_{\rm F}^4}=\frac{\mu}{\varepsilon_{\rm F}}-\frac{E}{E_{\rm FG}}=\frac{1}{2}\left(\frac{P}{P_{\rm FG}}-\frac{E}{E_{\rm FG}}\right).
\end{equation}
Using the mean-field result $C_{\rm MF}/k_{\rm F}^4=\alpha/2$, we can show that
\begin{equation}
\frac{C-C_{\rm MF}}{k_{\rm F}^4}=\nu-R.
\end{equation}
In Fig. \ref{fig6}, we show the quantity $(C-C_{\rm MF})/k_{\rm F}^4$ in the BCS-BEC crossover. We find that this difference is quite small in the entire BCS-BEC crossover and is peaked around $\eta\simeq0.7$, which agrees with recent quantum Monte Carlo results~\cite{QMC2D1,QMC2D2}.

\section{Summary}\label{s5}

The lack of a weakly interacting Bose condensate in the strong attraction limit is a longstanding problem for the theory of BCS-BEC crossover in two-dimensional Fermi gases. Especially, the mean-field prediction for the pressure in the BCS-BEC crossover shows the inadequacy of the
mean-field theory in 2D. The inadequacy of the 2D mean-field theory can be understood from the fact that the Born approximation for four-body scattering in 2D predicts an incorrect form of the composite boson coupling. In this work, we showed that this problem can be solved by including the contributions from the Gaussian quantum fluctuations. In the BEC limit, the missing logarithmic dependence on the boson chemical potential and hence the boson-boson interaction is naturally recovered by the quantum fluctuations. We determined the composite boson scattering length as
$a_{\rm B}\simeq 0.56 a_{\rm 2D}$, in good agreement with the exact four-body calculation and recent quantum Monte Carlo results. We calculated the chemical potential, the energy density, the pressure, and the contact for a homogeneous 2D Fermi gas in the BCS-BEC crossover. Our theoretical predictions are in good agreements with recent quantum Monte Carlo results and experimental measurements.

In the future, it is necessary to consider more many-body effects to explain the slight discrepancy between our theoretical prediction and the quantum Monte Carlo results, such as the GMB effect and the non-Gaussian fluctuations. In the BEC limit, an exact low-density expansion for the composite bosons~\cite{BEClimit} could also exist in 2D. It is also interesting to extend the present theoretical approach to the finite-temperature case and the spin-imbalanced case. The inclusion of the Gaussian fluctuations may provide better predictions for the Berezinskii-Kosterlitz-Thouless transition in the 2D BCS-BEC crossover~\cite{2Dexp8} and the phase structure of spin-imbalanced 2D Fermi gases~\cite{2Dexp6}.

{\bf Acknowledgments:} We thank Shiwei Zhang and Hao Shi for helpful discussions and Andrey Turlapov for useful communications. The work of Lianyi He was supported by the U. S. Department of Energy Nuclear Physics Office (Contract No. DE-AC02-05CH11231).  Haifeng L\"u was supported by NSFC
(Grant No. 61474018). Gaoqing Cao acknowledges the support from NSFC (Grant No. 11335005) and MOST (Grant No. 2013CB922000 and No. 2014CB845400).
Hui Hu and Xia-Ji Liu were supported by the ARC Discovery Projects (Grant No. FT130100815, No. FT140100003, No. DP140103231, and No. DP140100637).

\vspace{0.05in}
\appendix

\begin{widetext}
\section{Counting the infrared divergence of the function $f(\zeta)$}

Using the Taylor expansion $\ln(1-a)=-\sum_{n=1}^\infty a^n/n$, we can express the function $f(\zeta)$ as
\begin{eqnarray}\label{FEXP}
f(\zeta)&=&\frac{4}{\pi}\int_0^\infty dx\int_0^\infty dy\frac{{\cal A}{\cal C}+y^2{\cal B}{\cal D}+2{\cal F}^2}{{\cal A}^2+y^2{\cal B}^2}
-\frac{2\zeta^2}{\pi}\int_0^\infty dx\int_0^\infty dy\frac{{\cal C}^2+y^2{\cal D}^2}{{\cal A}^2+y^2{\cal B}^2} \nonumber\\
&&+\frac{2}{\pi}\sum_{n=2}^\infty\frac{\zeta^{2n-2}}{n}\sum_{k=0}^n\left(\begin{array}{cc}n \\ k\end{array}\right)2^{k}(-1)^{n-k}\zeta^{2n-2k}
\int_0^\infty dx\int_0^\infty dy\frac{({\cal A}{\cal C}+y^2{\cal B}{\cal D}+2{\cal F}^2)^k({\cal C}^2+y^2{\cal D}^2)^{n-k}}
{({\cal A}^2+y^2{\cal B}^2)^n}
\end{eqnarray}
To analyze the infrared divergence for $\zeta\rightarrow0$, we expand the quantities ${\cal A}$ and ${\cal B}$ in the denominators in powers
of $\zeta$,
\begin{equation}
{\cal A}(x,y)={\cal A}_0(x,y)+\sum_{n=1}^\infty\frac{\zeta^n}{n!}{\cal A}_n(x,y),\ \ \ \ \ \ \ \ \
{\cal B}(x,y)={\cal B}_0(x,y)+\sum_{n=1}^\infty\frac{\zeta^n}{n!}{\cal B}_n(x,y),
\end{equation}
where ${\cal A}_n=\partial^n{\cal A}/\partial\zeta^n|_{\zeta=0}$ and ${\cal B}_n=\partial^n{\cal B}/\partial\zeta^n|_{\zeta=0}$. The expansion coefficients ${\cal A}_n$ and ${\cal B}_n$ can be evaluated to arbitrary order by using Mathematica. Here we list the results for
${\cal A}_1$ and ${\cal B}_1$. We have
\begin{eqnarray}
&&{\cal A}_1=\int_0^\infty dz\frac{1}{L^2+y^2}\left[1+\frac{8xz}{\left(L^2-4xz\right)^{3/2}}\right]+\int_0^\infty dz\frac{2L^2}{\left(L^2+y^2\right)^2}\left(\frac{2}{\sqrt{L^2-4xz}}-1\right),\nonumber\\
&&{\cal B}_1=-\int_0^\infty dz\frac{2L}{L^2+y^2}\frac{1}{\left(L^2-4xz\right)^{3/2}}+\int_0^\infty dz\frac{2L}{\left(L^2+y^2\right)^2}\left(1-\frac{2}{\sqrt{L^2-4xz}}\right),
\end{eqnarray}
where $L\equiv z+1+x$ as defined in the text. In the infrared limit $x,y\rightarrow0$, we have ${\cal A}_1\rightarrow1$ and
${\cal B}_1\rightarrow-1$. The expansion of the quantity ${\cal A}^2+y^2{\cal B}^2$ takes the form
\begin{equation}\label{ABEXP}
{\cal A}^2(x,y)+y^2{\cal B}^2(x,y)={\cal A}_0^2(x,y)+y^2{\cal B}_0^2(x,y)+\sum_{n=1}^\infty \zeta^n\sum_{k=0}^n\frac{1}{k!(n-k)!}
\left[{\cal A}_k(x,y){\cal A}_{n-k}(x,y)+y^2{\cal B}_k(x,y){\cal B}_{n-k}(x,y)\right].
\end{equation}
For further analysis, it is convenient to use the polar coordinates $x=\rho\cos\varphi$ and $y=\rho\sin\varphi$. At exactly $\zeta=0$, we have
${\cal A}_0^2+y^2{\cal B}_0^2\simeq\rho^2$ in the infrared limit $\rho\rightarrow0$. To capture the leading asymptotic behavior, we find that it is sufficient to approximate the quantity ${\cal A}^2+y^2{\cal B}^2$ as
\begin{equation}\label{EXPAPP}
{\cal A}^2(x,y)+y^2{\cal B}^2(x,y)\simeq {\cal J}(x,y)={\cal A}_0^2(x,y)+y^2{\cal B}_0^2(x,y)+2\zeta{\cal A}_0(x,y){\cal A}_1(x,y)+\zeta^2{\cal A}_1^2(x,y).
\end{equation}
In the infrared limit $\rho\rightarrow0$, the function ${\cal J}(x,y)$ behaves as
\begin{equation}
{\cal J}(x,y)\simeq\rho^2+2\zeta\rho\cos\varphi+\zeta^2.
\end{equation}
The other contributions we neglected in approximation (\ref{EXPAPP}) behave in the infrared limit as
\begin{equation}
\left(\sum_{n=1}^\infty a_n\zeta^n\right)\rho^2+\left(\sum_{n=2}^\infty b_n\zeta^n\right)\rho+\left(\sum_{n=3}^\infty c_n\zeta^n\right).
\end{equation}
The above terms lead to vanishing contributions in the limit $\zeta\rightarrow0$. One can prove this observation by carefully analyzing the infrared behavior of the following integral
\begin{equation}
I_{mn}=\int_0^\epsilon \rho d\rho\frac{\rho^m}{(\rho^2+2\zeta\rho\cos\varphi+\zeta^2)^n}.
\end{equation}
The properties of the integral $I_{mn}$ can be summarized as follows: For $m>2(n-1)$, the integral is finite; for $m=2(n-1)$, it diverges as $I_{mn}\sim-\ln\zeta$; and for $m<2(n-1)$, we have $I_{mn}\sim1/\zeta^{2n-2-m}$ for $\zeta\rightarrow0$.

\begin{figure}[!htb]
\begin{center}
\includegraphics[width=9cm]{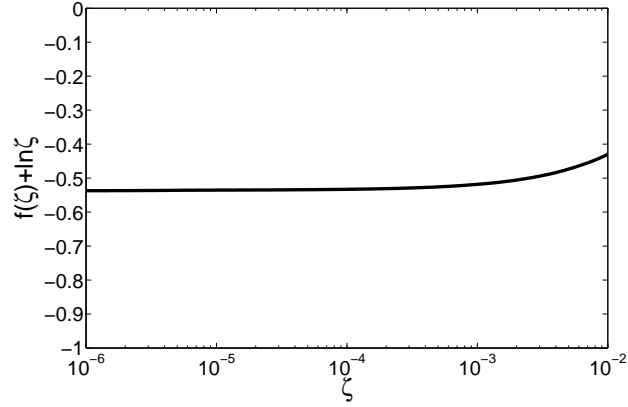}
\caption{ The quantity $f(\zeta)+\ln(\zeta)$ as a function of $\zeta$ in the range $10^{-6}<\zeta<10^{-2}$. In the calculation we use Eq. (A16) for $f(\zeta)$. \label{fig7}}
\end{center}
\end{figure}

In the infrared limit $\rho\rightarrow0$, the second term in the expansion, (\ref{FEXP}), behaves as
\begin{equation}
-\frac{2\zeta^2}{\pi}\int_0^\infty dx\int_0^\infty dy\frac{{\cal C}^2+y^2{\cal D}^2}{{\cal A}^2+y^2{\cal B}^2} \sim
-\frac{2\zeta^2}{\pi}\int_0^{\pi/2}d\varphi\int_0^\epsilon\rho d\rho
\frac{\frac{1}{16}+\frac{1}{25}\rho^2\sin^2\varphi}{\rho^2+2\zeta\rho\cos\varphi+\zeta^2}.
\end{equation}
It vanishes in the limit $\zeta\rightarrow0$. The third term in the expansion, (\ref{FEXP}), behaves as
\begin{eqnarray}
&&\frac{2}{\pi}\sum_{n=2}^\infty\frac{\zeta^{2n-2}}{n}\sum_{k=0}^n\left(\begin{array}{cc}n \\ k\end{array}\right)2^{k}(-1)^{n-k}\zeta^{2n-2k}
\int_0^\infty dx\int_0^\infty dy\frac{({\cal A}{\cal C}+y^2{\cal B}{\cal D}+2{\cal F}^2)^k({\cal C}^2+y^2{\cal D}^2)^{n-k}}
{({\cal A}^2+y^2{\cal B}^2)^n}\nonumber\\
&\sim&\frac{2}{\pi}\sum_{n=2}^\infty\frac{\zeta^{2n-2}}{n}\sum_{k=0}^n\left(\begin{array}{cc}n \\ k\end{array}\right)2^{k}(-1)^{n-k}\zeta^{2n-2k}
\int_0^{\pi/2}d\varphi\int_0^\epsilon\rho d\rho\frac{(\rho\cos\varphi/4+\rho^2\sin^2\varphi/5+1/2)^k(1/16+\rho^2\sin^2\varphi/25)^{n-k}}
{(\rho^2+2\zeta\rho\cos\varphi+\zeta^2)^n}.
\end{eqnarray}
A careful analysis shows that this term leads to a finite contribution in the limit $\zeta\rightarrow0$. The nonvanishing contribution from the $k=n$ terms can be expressed as
\begin{eqnarray}
&&\frac{2}{\pi\zeta^2}\sum_{n=2}^\infty\frac{1}{n}\int_0^\infty dx\int_0^\infty dy\left(\frac{4\zeta^2{\cal F}^2}{{\cal A}^2+y^2{\cal B}^2}\right)^n\nonumber\\
&=&-\frac{2}{\pi\zeta^2}\int_0^\infty dx\int_0^\infty dy\ln\left(1-4\zeta^2\frac{{\cal F}^2}{{\cal A}^2+y^2{\cal B}^2}\right)
-\frac{8}{\pi}\int_0^\infty dx\int_0^\infty dy\frac{{\cal F}^2}{{\cal A}^2+y^2{\cal B}^2}.
\end{eqnarray}
Next we analyze the first term in (\ref{FEXP}), which develops the logarithmic divergence. In the infrared limit, it behaves as
\begin{equation}
\frac{4}{\pi}\int_0^\infty dx\int_0^\infty dy\frac{{\cal A}{\cal C}+y^2{\cal B}{\cal D}+2{\cal F}^2}{{\cal A}^2+y^2{\cal B}^2}
\sim\frac{4}{\pi}\int_0^{\pi/2}d\varphi\int_0^\epsilon\rho d\rho
\frac{\rho\cos\varphi/4+\rho^2\sin^2\varphi/5+1/2}{\rho^2+2\zeta\rho\cos\varphi+\zeta^2}.
\end{equation}
Therefore, we can separate the above contribution into two pieces. The finite piece is given by
\begin{equation}
\frac{4}{\pi}\int_0^\infty dx\int_0^\infty dy\frac{{\cal A}{\cal C}+y^2{\cal B}{\cal D}}{{\cal A}^2+y^2{\cal B}^2}.
\end{equation}
The divergent piece is given by
\begin{equation}
\frac{8}{\pi}\int_0^\infty dx\int_0^\infty dy\frac{{\cal F}^2}{{\cal A}^2+y^2{\cal B}^2}.
\end{equation}
In the infrared limit, this piece behaves as
\begin{equation}
\frac{8}{\pi}\int_0^{\pi/2}d\varphi\int_0^\epsilon\rho d\rho
\frac{1/4}{\rho^2+2\zeta\rho\cos\varphi+\zeta^2}\sim-\ln\zeta.
\end{equation}
Therefore, it exactly develops the asymptotic behavior $f(\zeta)\sim-\ln\zeta+\lambda$ for $\zeta\rightarrow0$. Summarizing the nonvanishing pieces (A11), (A13), and (A14), we find that to capture the logarithmic divergence and determine the finite term $\lambda$, it is sufficient to approximate the function $f(\zeta)$ as
\begin{equation}
f(\zeta)\simeq-\frac{2}{\pi\zeta^2}\int_0^\infty dx\int_0^\infty dy\ln\left[1-4\zeta^2\frac{{\cal F}_0^2(x,y)}{{\cal J}(x,y)}\right]
+\frac{4}{\pi}\int_0^\infty dx\int_0^\infty dy\frac{{\cal A}_0(x,y){\cal C}_0(x,y)+y^2{\cal B}_0(x,y){\cal D}_0(x,y)}
{{\cal A}_0^2(x,y)+y^2{\cal B}_0^2(x,y)}.
\end{equation}
In Fig. \ref{fig7} we show the numerical result of $f(\zeta)+\ln(\zeta)$ in the range $10^{-6}<\zeta<10^{-2}$. It is clear that in the limit $\zeta\rightarrow0$, it converges to a constant. Thus we determine $\lambda\simeq-0.54$.

\end{widetext}


\begin{thebibliography}{99}
\bibitem{Eagles}      {D. M. Eagles, \emph{Possible pairing without superconductivity at low carrier concentrations in bulk and thin-film
                       superconducting semiconductors}, Phys. Rev. {\bf 186}, 456 (1969).}
\bibitem{Leggett}     {A. J. Leggett, \emph{Diatomic molecules and Cooper pairs}, in \emph{Modern Trends in the Theory of Condensed Matter},
                       \emph{Lecture Notes in Physics}, Vol. {\bf 115} (Springer-Verlag, Berlin, 1980).}
\bibitem{NSR}         {P. Nozieres and S. Schmitt-Rink, \emph{Bose condensation in an attractive fermion gas: From weak to strong coupling
                       superconductivity}, J. Low Temp. Phys. {\bf 59}, 195 (1985).}
\bibitem{BCSBEC1}     {C. A. R. Sa de Melo, M. Randeria, and J. R. Engelbrecht, \emph{Crossover from BCS to Bose superconductivity:
                       Transition temperature and time-dependent Ginzburg-Landau theory}, Phys. Rev. Lett. {\bf 71}, 3202 (1993).}
\bibitem{BCSBEC2}     {J. R. Engelbrecht, M. Randeria, and C. A. R. Sa de Melo, \emph{BCS to Bose crossover: Broken-symmetry state},
                       Phys. Rev. {\bf B55}, 15153 (1997).}
\bibitem{BCSBEC3}     {Q. Chen, J. Stajic, S. Tan, and K. Levin, \emph{BCS-BEC crossover: From high temperature superconductors to ultracold
                       superfluids}, Phys. Rep. {\bf 412}, 1 (2005).}
\bibitem{BCSBEC4}     {V. Gurarie, and L. Radzihovsky, \emph{Resonantly-paired fermionic superfluids},  Ann. Phys. (N. Y.) {\bf 322}, 2 (2007).}
\bibitem{BCSBEC5}     {S. Giorgini, L. P. Pitaevskii, and S. Stringari, \emph{Theory of ultracold atomic Fermi gases},
                       Rev. Mod. Phys. {\bf 80}, 1215 (2008).}
\bibitem{BCSBECexp1}  {M. Greiner, C. A. Regal, and D. S. Jin, \emph{Emergence of a molecular Bose-Einstein condensate from a Fermi gas},
                       Nature {\bf 426}, 537 (2003).}
\bibitem{BCSBECexp2}  {S. Jochim, M. Bartenstein, A. Altmeyer, G. Hendl, S. Riedl, C. Chin, J. Hecker Denschlag, and R. Grimm,
                       \emph{Bose-Einstein condensation of molecules}, Science {\bf 302}, 2101(2003).}
\bibitem{BCSBECexp3}  {M. W. Zwierlein, J. R. Abo-Shaeer, A. Schirotzek, C. H. Schunck, and W. Ketterle,
                       \emph{Vortices and superfluidity in a strongly interacting Fermi gas}, Nature {\bf 435}, 1047 (2003).}
\bibitem{FR1}         {T. Koehler, K. Goral, and P. S. Julienne, \emph{Production of cold molecules via magnetically tunable Feshbach resonances},
                       Rev. Mod. Phys. {\bf 78}, 1311 (2006).}
\bibitem{FR2}         {C. Chin, R. Grimm, P. Julienne, and E. Tiesinga, \emph{Feshbach resonances in ultracold gases},
                       Rev. Mod. Phys. {\bf 82}, 1225 (2010).}
\bibitem{TH01}        {A. Perali, P. Pieri, L. Pisani, and G. C. Strinati, \emph{BCS-BEC crossover at finite temperature for superfluid trapped
                       Fermi atoms}, Phys. Rev. Lett. {\bf 92}, 220404 (2004).}
\bibitem{TH02}        {P. Pieri, L. Pisani, and G. C. Strinati, \emph{BCS-BEC crossover at finite temperature in the broken-symmetry phase},
                       Phys. Rev. {\bf B70}, 094508 (2004).}
\bibitem{TH03}        {H. Hu, X.-J. Liu, and P. D. Drummond, \emph{Equation of state of a superfluid Fermi gas in the BCS-BEC crossover},
                       Europhys. Lett. {\bf 74}, 574 (2006).}
\bibitem{TH04}        {R. B. Diener, R. Sensarma, and M. Randeria, \emph{Quantum fluctuations in the superfluid state of the BCS-BEC crossover},
                       Phys. Rev. {\bf A77}, 023626 (2008).}
\bibitem{TH05}        {H. Hu, X. -J. Liu, and P. D. Drumond, \emph{Universal thermodynamics of strongly interacting Fermi gases},
                       Nat. Phys. {\bf 3}, 469 (2007).}
\bibitem{TH06}        {Y. Nishida and D. T. Son, \emph{$\epsilon$ expansion for a Fermi gas at infinite scattering length},
                       Phys. Rev. Lett. {\bf 97}, 050403 (2006).}
\bibitem{TH07}        {R. Haussmann, W. Rantner, S. Cerrito, and W. Zwerger, \emph{Thermodynamics of the BCS-BEC crossover},
                       Phys. Rev. {\bf A75}, 023610 (2007).}
\bibitem{TH08}        {M. Y. Veillette, D. E. Sheehy, and L. Radzihovsky, \emph{Large-$N$ expansion for unitary superfluid Fermi gases},
                       Phys. Rev. {\bf A75}, 043614 (2007).}
\bibitem{TH09}        {Y. Ohashi and A. Griffin, \emph{Superfluidity and collective modes in a uniform gas of Fermi atoms with a Feshbach resonance},
                       Phys. Rev. {\bf A67}, 063612 (2003).}
\bibitem{TH10}        {E. Taylor, A. Griffin, N. Fukushima, and Y. Ohashi, \emph{Pairing fluctuations and the superfluid density through the BCS-BEC
                       crossover}, Phys. Rev. {\bf A74}, 063626 (2006).}
\bibitem{TH11}        {N. Fukushima, Y. Ohashi, E. Taylor, and A. Griffin, \emph{Superfluid density and condensate fraction in the BCS-BEC crossover
                       regime at finite temperatures}, Phys. Rev. {\bf A75}, 033609 (2007).}
\bibitem{EOSexp1}     {S. Nascimbene, N. Navon, K. Jiang, F. Chevy, and C. Salomon, \emph{Exploring the thermodynamics of a universal Fermi gas},
                       Nature {\bf 463}, 1057 (2010).}
\bibitem{EOSexp2}     {N. Navon, S. Nascimbene, F. Chevy, and C. Salomon,  \emph{The equation of state of a low-temperature Fermi gas with tunable
                       interactions}, Science {\bf 328}, 729 (2010).}
\bibitem{EOSexp3}     {M. J. H. Ku, A. T. Sommer, L. W. Cheuk, and M. W. Zwierlein, \emph{Revealing the superfluid lambda transition in the universal
                       thermodynamics of a unitary Fermi gas}, Science {\bf 335}, 563 (2012).}
\bibitem{EOSmc1}      {J. Carlson, S.-Y. Chang, V. R. Pandharipande, and K. E. Schmidt, \emph{Superfluid Fermi gases with large scattering length},
                       Phys. Rev. Lett. {\bf 91}, 050401 (2003).}
\bibitem{EOSmc2}      {C. Lobo, A. Recati, S. Giorgini, and S. Stringari, \emph{Normal state of a polarized Fermi gas at unitarity},
                       Phys. Rev. Lett. {\bf 97}, 200403 (2006).}
\bibitem{EOSmc3}      {M. M. Forbes, S. Gandolfi, and A. Gezerlis, \emph{Resonantly interacting fermions in a box},
                       Phys. Rev. Lett. {\bf 106}, 235303 (2011).}
\bibitem{EOSmc4}      {J. Carlson, S. Gandolfi, K. E. Schmidt, and S. Zhang, \emph{Auxiliary-field quantum Monte Carlo method for strongly paired
                       fermions}, Phys. Rev. {\bf A84}, 061602(R) (2011).}
\bibitem{BCSBEC2D-1}  {K. Miyake, \emph{Fermi liquid theory of dilute submonolayer $^3$He on thin $^4$He II film: Dimer bound state and
                       cooper pairs}, Prog. Theor. Phys. {\bf 69}, 1794 (1983).}
\bibitem{BCSBEC2D-2}  {M. Randeria, J.-M. Duan, and L.-Y. Shieh, \emph{Bound states, Cooper pairing, and Bose condensation in two dimensions},
                       Phys. Rev. Lett. {\bf 62}, 981 (1989).}
\bibitem{BCSBEC2D-3}  {M. Randeria, J.-M. Duan, and L.-Y. Shieh, \emph{Superconductivity in a two-dimensional Fermi gas: Evolution from Cooper
                       pairing to Bose condensation}, Phys. Rev. {\bf B41}, 327 (1990). }
\bibitem{BCSBEC2D-4}  {V. M. Loktev, R. M. Quick, and S. G. Sharapov, \emph{Phase fluctuations and pseudogap phenomena},
                       Phys. Rep. {\bf 349}, 1 (2001).}
\bibitem{2Dexp1}  {K. Martiyanov, V. Makhalov, and A. Turlapov, \emph{Observation of a two-dimensional Fermi gas of atoms},
                   Phys. Rev. Lett. {\bf 105}, 030404 (2010).}
\bibitem{2Dexp2}  {V. Makhalov, K. Martiyanov, and A. Turlapov, \emph{Ground-state pressure of quasi-2D Fermi and Bose gases},
                   Phys. Rev. Lett. {\bf 112}, 045301 (2014).}
\bibitem{2Dexp3}  {B. Fr{\"o}hlich, M. Feld, E. Vogt, M. Koschorreck, W. Zwerger, and M. K{\"o}hl,
                   \emph{Radiofrequency spectroscopy of a strongly interacting two-dimensional Fermi gas},
                   Phys. Rev. Lett. {\bf 106}, 105301 (2011).}
\bibitem{2Dexp4}  {M. Feld, B. Fr{\"o}hlich, E. Vogt, M. Koschorreck, and M K{\"o}hl,
                   \emph{Observation of a pairing pseudogap in a two-dimensional Fermi gas}, Nature {\bf 480}, 75 (2011).}
\bibitem{2Dexp5}  {Y. Zhang, W. Ong, I. Arakelyan, and J. E. Thomas,
                   \emph{Polarons in the radio-frequency spectrum of a quasi-two-dimensional Fermi gas},
                   Phys. Rev. Lett. {\bf 108}, 235302 (2012).}
\bibitem{2Dexp6}  {W. Ong, C.-Y. Cheng, I. Arakelyan, and J. E. Thomas,
                   \emph{Spin-imbalanced quasi-two-dimensional Fermi gases},
                   Phys. Rev. Lett. {\bf 114}, 110403 (2015).}
\bibitem{2Dexp7}  {M. G. Ries, A. N. Wenz, G. Z¨¹rn, L. Bayha, I. Boettcher, D. Kedar, P. A. Murthy, M. Neidig, T. Lompe, and S. Jochim,
                   \emph{Observation of pair condensation in the quasi-2D BEC-BCS crossover}, Phys. Rev. Lett. {\bf 114}, 230401 (2015).}
\bibitem{2Dexp8}  {P. A. Murthy, I. Boettcher, L. Bayha, M. Holzmann, D. Kedar, M. Neidig, M. G. Ries, A. N. Wenz, G. Z¨¹rn, and S. Jochim,
                   \emph{Observation of the Berezinskii-Kosterlitz-Thouless phase transition in an ultracold Fermi gas}, arXiv:1505.02123.}
\bibitem{2Dexp9}   {A. T. Sommer, L. W. Cheuk, M. J. H. Ku, W. S. Bakr, and M. W. Zwierlein,
                    \emph{Evolution of fermion pairing from three to two dimensions}, Phys. Rev. Lett. {\bf 108}, 045302 (2012).}
\bibitem{2Dexp10}  {P. Dyke, E. D. Kuhnle, S. Whitlock, H. Hu, M. Mark, S. Hoinka, M. Lingham, P. Hannaford, and C. J. Vale,
                    \emph{Crossover from 2D to 3D in a weakly interacting Fermi gas}, Phys. Rev. Lett. {\bf 106}, 105304 (2011).}
\bibitem{QMC2D1}  {G. Bertaina and S. Giorgini, \emph{BCS-BEC crossover in a two-dimensional Fermi gas},  Phys. Rev. Lett. {\bf 106}, 110403 (2011).}
\bibitem{QMC2D2}  {H. Shi, S. Chiesa, and S. Zhang, \emph{Exact ground-state properties of strongly interacting Fermi gases in two dimensions},
                   arXiv:1504.00925.}
\bibitem{QMC2D3}  {E. R. Anderson and J. E. Drut, Pressure, \emph{compressibility, and contact of the two-dimensional attractive Fermi gas},
                   arXiv:1505.01525.}
\bibitem{2DTH01}  {D. S. Petrov, M. A. Baranov, and G. V. Shlyapnikov, \emph{Superfluid transition in quasi-two-dimensional Fermi gases},
                   Phys. Rev. {\bf A67}, 031601(R) (2003).}
\bibitem{2DTH02}    {J.-P. Martikainen and P. Torma, \emph{Quasi-two-dimensional superfluid fermionic gases},
                     Phys. Rev. Lett. {\bf 95}, 170407 (2005).}
\bibitem{2DTH03}    {J. Tempere, M. Wouters and J. T. Devreese, \emph{Imbalanced Fermi superfluid in a one-dimensional optical potential},
                     Phys. Rev. {\bf B75}, 184526 (2007).}
\bibitem{2DTH04}    {W. Zhang, G.-D. Lin, and L.-M. Duan, \emph{Berezinskii-Kosterlitz-Thouless transition in a trapped quasi-two-dimensional Fermi
                     gas near a Feshbach resonance}, Phys. Rev. {\bf A78}, 043617 (2008).}
\bibitem{2DTH05}    {G. J. Conduit, P. H. Conlon and B. D. Simons, \emph{Superfluidity at the BEC-BCS crossover in two-dimensional Fermi gases with
                     population and mass imbalance}, Phys. Rev. {\bf A77}, 053617 (2008).}
\bibitem{2DTH06}    {L. He and P. Zhuang, \emph{Phase diagram of a cold polarized Fermi gas in two dimensions}, Phys. Rev. {\bf A78}, 033613 (2008).}
\bibitem{2DTH07}  {M. Iskin and C. A. R. Sa de Melo, \emph{Evolution from BCS to Berezinskii-Kosterlitz-Thouless superfluidity in one-dimensional
                   optical lattices}, Phys. Rev. Lett. {\bf 103}, 165301 (2009).}
\bibitem{2DTH08}  {A. A. Orel, P. Dyke, M. Delehaye, C. J. Vale, and H. Hu, \emph{Density distribution of a trapped two-dimensional strongly
                   interacting Fermi gas},  New J. Phys. {\bf 13}, 113032 (2011).}
\bibitem{2DTHs}    {S. Zoellner, G. M. Bruun, and C. J. Pethick, \emph{Polarons and molecules in a two-dimensional Fermi gas},
                     Phys. Rev. {\bf A83}, 021603(R) (2011).}
\bibitem{2DTH09}    {V. Pietila, \emph{Pairing and radio-frequency spectroscopy in two-dimensional Fermi gases},
                     Phys. Rev. {\bf A86}, 023608 (2012).}
\bibitem{2DTH10}    {E. Taylor and M. Randeria, \emph{Apparent low-energy scale invariance in two-dimensional Fermi gases},
                     Phys. Rev. Lett. {\bf 109}, 135301 (2012).}
\bibitem{2DTH11}    {J. Hofmann, \emph{Quantum anomaly, universal relations, and breathing mode of a two-dimensional Fermi gas},
                     Phys. Rev. Lett. {\bf 108}, 185303 (2012).}
\bibitem{2DTH12}    {L. He and X.-G. Huang, \emph{BCS-BEC crossover in 2D Fermi gases with Rashba spin-orbit coupling},
                     Phys. Rev. Lett. {\bf 108}, 145302 (2012).}
\bibitem{2DTH13}    {H. Caldas, A. L. Mota, R. L. S. Farias, and L. A. Souza, \emph{Superfluidity in two-dimensional imbalanced Fermi gases},
                     J. Stat. Mech.: Theory Exp. P10019 (2012).}
\bibitem{2DTH14}    {M. A. Resende, A. L. Mota, R. L. S. Farias, and H. Caldas, \emph{Finite temperature phase diagram of quasi-two-dimensional
                     imbalanced Fermi gases beyond mean-field}, Phys. Rev. {\bf A86}, 033603 (2012).}
\bibitem{2DTH15}    {S. N. Klimin, J. Tempere, and J. T. Devreese, \emph{Pseudogap and preformed pairs in the imbalanced Fermi gas in two
                     dimensions}, New J. Phys. {\bf 14}, 103044 (2012).}
\bibitem{2DTH16}    {V. Ngampruetikorn, J. Levinsen, and M. M. Parish, \emph{Pair correlations in the two-dimensional Fermi gas},
                     Phys. Rev. Lett. {\bf 111}, 265301 (2013).}
\bibitem{2DTH17}    {M. M. Parish and J. Levinsen, \emph{Highly polarized Fermi gases in two dimensions}, Phys. Rev. {\bf A87}, 033616 (2013).}
\bibitem{2DTH18}    {L. Salasnich, P. A. Marchetti, and F. Toigo, \emph{Superfluidity, sound velocity, and quasicondensation in the two-dimensional
                     BCS-BEC crossover}, Phys. Rev. {\bf A88}, 053612 (2013).}
\bibitem{2DTH19}    {A. M. Fischer and M. M. Parish, \emph{BCS-BEC crossover in a quasi-two-dimensional Fermi gas},
                     Phys. Rev. {\bf A88}, 023612 (2013).}
\bibitem{2DTH20}    {P. Strack and P. Jakubczyk, \emph{Fluctuations of imbalanced fermionic superfluids in two dimensions induce continuous quantum
                     phase transitions and non-Fermi-liquid behavior}, Phys. Rev. {\bf X4}, 021012 (2014).}
\bibitem{2DTH21}    {M. Bauer, M. M. Parish, and T. Enss, \emph{Universal equation of state and pseudogap in the two-dimensional Fermi gas},
                     Phys. Rev. Lett. {\bf 112}, 135302 (2014).}
\bibitem{2DTH22}    {M. Barth and J. Hofmann, \emph{Pairing effects in the nondegenerate limit of the two-dimensional Fermi gas},
                     Phys. Rev. {\bf A89}, 013614 (2014).}
\bibitem{2DTH23}    {F. Marsiglio, P. Pieri, A. Perali, F. Palestini, and G. C. Strinati, \emph{Pairing effects in the normal phase of a
                     two-dimensional Fermi gas}, Phys. Rev. {\bf B91}, 054509 (2015).}
\bibitem{2DTH24}    {L. Salasnich and F. Toigo, \emph{Composite bosons in the two-dimensional BCS-BEC crossover from Gaussian fluctuations},
                     Phys. Rev. {\bf A91}, 011604(R) (2015).}
\bibitem{2DTH25}    {D. E. Sheehy, \emph{Fulde-Ferrell-Larkin-Ovchinnikov state of two-dimensional imbalanced Fermi gases}, arXiv: 1407.8021.}
\bibitem{FB3D}    {D. S. Petrov, C. Salomon, and G. V. Shlyapnikov, \emph{Weakly bound dimers of fermionic atoms},
                   Phys. Rev. Lett. {\bf 93}, 090404 (2004).}
\bibitem{2DBEC01}   {V. N. Popov, \emph{On the theory of the superfluidity of two- and one-dimensional Bose systems},
                     Theor. Math. Phys. {\bf 11}, 565 (1972).}
\bibitem{2DBEC02}   {A. Y. Cherny and A. A. Shanenko, \emph{Dilute Bose gas in two dimensions: Density expansions and the Gross-Pitaevskii equation},
                     Phys. Rev. {\bf E64}, 027105 (2001).}
\bibitem{2DBEC03}   {C. Mora and Y. Castin, \emph{Extension of Bogoliubov theory to quasicondensates}, Phys. Rev. {\bf A67}, 053615 (2003).}
\bibitem{2DBEC04}   {L. Pricoupenko, \emph{Variational approach for the two-dimensional trapped Bose-Einstein condensate},
                     Phys. Rev. {\bf A70}, 013601 (2004).}
\bibitem{2DBEC05}   {G. E. Astrakharchik, J. Boronat, J. Casulleras, I. L. Kurbakov, and Yu. E. Lozovik,
                     \emph{Equation of state of a weakly interacting two-dimensional Bose gas studied at zero temperature by means of
                     quantum Monte Carlo methods}, Phys. Rev. {\bf A79}, 051602(R) (2009).}
\bibitem{2DBEC06}   {C. Mora and Y. Castin, \emph{Ground state energy of the two-dimensional weakly interacting Bose gas:
                     first correction beyond Bogoliubov theory}, Phys. Rev. Lett. {\bf 102}, 180404 (2009).}
\bibitem{2DBEC07}   {S. R. Beane, \emph{Ground-state energy of the interacting Bose gas in two dimensions: An explicit construction},
                     Phys. Rev. {\bf A82}, 063610 (2010).}
\bibitem{PEOS2D1} {P. Bloom, \emph{Two-dimensional Fermi gas}, Phys. Rev. {\bf B12}, 125 (1975).}
\bibitem{PEOS2D2} {J. R. Engelbrecht, M. Randeria, and L. Zhang, \emph{Landau $f$ function for the dilute Fermi gas in two dimensions},
                   Phys. Rev. {\bf B45}, 10135(R) (1992).}
\bibitem{PEOS2D3} {L. He, \emph{Interaction energy and itinerant ferromagnetism in a strongly interacting Fermi gas in the absence of molecule
                   formation}, Phys. Rev. {\bf A90}, 053633 (2014).}
\bibitem{Blave1}  {T. D. Cohen, \emph{Functional integrals for QCD at nonzero chemical potential and zero density},
                   Phys. Rev. Lett. {\bf 91}, 222001 (2003).}
\bibitem{Blave2}  {L. He, \emph{Nambu--Jona-Lasinio model description of weakly interacting Bose condensate and BEC-BCS crossover in dense QCD-like
                   theories}, Phys. Rev. {\bf D82}, 096003 (2010). }
\bibitem{N2D}     {J. Keeling, P. R. Eastham, M. H. Szymanska, and P. B. Littlewood, \emph{BCS-BEC crossover in a system of microcavity polaritons},
                   Phys. Rev. {\bf B72}, 115320 (2005).}
\bibitem{Coupling2D} {M. Drechsler and W. Zwerger, \emph{Crossover from BCS-superconductivity to Bose-condensation},
                      Annalen der Physik {\bf 1}, 15 (1992).}
\bibitem{GPFhe}   {L. He and X.-G. Huang, \emph{BCS-BEC crossover in three-dimensional Fermi gases with spherical spin-orbit coupling},
                   Phys. Rev. {\bf B86}, 014511 (2012).}
\bibitem{GPE2D}   {M. D. Lee, S. A. Morgan, M. J. Davis, and K. Burnett, \emph{Energy dependent scattering and the Gross-Pitaevskii equation in two
                   dimensional Bose-Einstein condensates}, Phys. Rev. {\bf A65}, 043617 (2002).}
\bibitem{FourB}   {I. V. Brodsky, M. Yu. Kagan, A. V. Klaptsov, R. Combescot, and X. Leyronas, \emph{Exact diagrammatic approach for dimer-dimer
                   scattering and bound states of three and four resonantly interacting particles}, Phys. Rev. {\bf A73}, 032724 (2006).}
\bibitem{GMB}     {L. P. Gor'kov and T. K. Melik-Barkhudarov, \emph{Contribution to the theory of superfluidity in an imperfect fermi gas},
                   Sov. Phys. JETP {\bf 13}, 1018 (1961).}
\bibitem{Contact01} {S. Tan, \emph{Energetics of a strongly correlated Fermi gas}, Ann. Phys. (N. Y.), {\bf 323}, 2952 (2008).}
\bibitem{Contact02} {S. Tan, \emph{Large momentum part of fermions with large scattering length}, Ann. Phys. (N. Y.) {\bf 232}, 2971 (2008).}
\bibitem{Contact2D} {F. Werner and Y. Castin, \emph{General relations for quantum gases in two and three dimensions: Two-component fermions},
                     Phys. Rev. {\bf A86}, 013626 (2012).}
\bibitem{BEClimit}  {X. Leyronas and R. Combescot, \emph{Superfluid equation of state of dilute composite bosons},
                     Phys. Rev. Lett. {\bf 99}, 170402 (2007).}
\end{thebibliography}
\end{document}